\def\E{\end{document}}
\documentclass[11pt]{article}
\usepackage{amssymb,amsmath}

\begin{document}
\title{\bf
Morawetz estimates and spacetime bounds for quasilinear Schr\"{o}dinger equations with critical Sobolev exponent
}
  \author{Xianfa Song{\thanks{E-mail: songxianfa@tju.edu.cn(X.F. Song)
 }} \\
\small Department of Mathematics, School of Mathematics, Tianjin University,\\
\small Tianjin, 300072, P. R. China
}

\maketitle
\date{}

\newtheorem{theorem}{Theorem}[section]
\newtheorem{definition}{Definition}[section]
\newtheorem{lemma}{Lemma}[section]
\newtheorem{proposition}{Proposition}[section]
\newtheorem{corollary}{Corollary}[section]
\newtheorem{remark}{Remark}[section]
\renewcommand{\theequation}{\thesection.\arabic{equation}}
\catcode`@=11 \@addtoreset{equation}{section} \catcode`@=12

\begin{abstract}

In this paper, we study the following Cauchy problem
\begin{equation*}
\left\{
\begin{array}{lll}
iu_t=\Delta u+2uh'(|u|^2)\Delta h(|u|^2)+F(|u|^2)u\mp A[h(|u|^2]^{2^*-1}h'(|u|^2)u,\ x\in \mathbb{R}^N, \ t>0\\
u(x,0)=u_0(x),\quad x\in \mathbb{R}^N.
\end{array}\right.
\end{equation*}
Here $h(s)$ and $F(s)$ are some real-valued functions, $h(s)\geq 0$ and $h'(s)\geq 0$ for $s\geq 0$, $N\geq 3$, $A>0$.
Besides obtaining sufficient conditions on the blowup in finite time and global existence of the solution, we establish Morawetz estimates and spacetime bounds for the global solution based on pseudoconformal conservation law, which is an important tool to construct scattering operator on the energy space.

{\bf Keywords:} Qusilinear Schr\"{o}dinger equation; Global existence; Blow up; Pseudoconformal conservation law; Morawetz estimate; Spacetime bound.

{\bf 2000 MSC: 35Q55.}

\end{abstract}

\section{Introduction}
\qquad In this paper, we consider the following Cauchy problem:
\begin{equation}
\label{1} \left\{
\begin{array}{lll}
iu_t=\Delta u+2uh'(|u|^2)\Delta h(|u|^2)+F(|u|^2)u\mp A[h(|u|^2]^{2^*-1}h'(|u|^2)u,\ x\in \mathbb{R}^N, \ t>0\\
u(x,0)=u_0(x),\quad x\in \mathbb{R}^N.
\end{array}\right.
\end{equation}
Here $N\geq 3$, $A>0$, $h(s)$ and $F(s)$ are some real-valued functions, $h(s)\geq 0$ and $h'(s)\geq 0$ for $s\geq 0$, and there exists positive constant $M$ such that
\begin{align}
 \frac{h(s)}{s^{\frac{1}{2}}}\geq M>0,\quad  {\rm including}\quad \frac{h(s)}{s^{\frac{1}{2}}}\rightarrow +\infty,\quad {\rm as} \ s\rightarrow +\infty.\label{3251}
\end{align}
(\ref{1}) often appears in condensed matter theory, in plasma physics and fluid mechanics and in the theory of Heisenberg ferromagnet and magnons, see \cite{Bass, Goldman, Litvak, Makhankov}. It can be used to illustrate many physical phenomena. For example, if $h(s)=s$, it is called the superfluid film equation in plasma physics or the modified nonlinear Schr\"odinger equation(\cite{Ku, LSS}); If $h(s)=\sqrt{1+s}$, it models the self-channelling of a high-power ultra short laser in matter while if $h(s)=\sqrt{s}$, it illustrates the physical phenomenon in dissipative quantum mechanics, see \cite{Borovskii, Bouard, Hasse, Ritchie}. The local well-posedness of the solution to (\ref{1}) has been established by many authors, see \cite{Colin1, Kenig, Poppenberg1} and the references therein. In convenience, we call (\ref{1}) with the term $-A[h(|u|^2]^{2^*-1}h'(|u|^2)u$ as (\ref{1}A) and (\ref{1}) with the term $A[h(|u|^2]^{2^*-1}h'(|u|^2)u$ as (\ref{1}B). An interesting topic on (\ref{1}) is the global existence and blowup phenomena. We state the precise definition of global existence and finite time blowup of solutions.

{\bf Definition 1.} {\it Let $u(x,t)$ be the solution of (\ref{1}). We say that $u(x,t)$ is global existence if the maximum existence interval of $u(x,t)$ for $t$ is $[0, +\infty)$. Otherwise, we say that $u(x,t)$ will blow up in finite time if there exists a time $0<T<+\infty$ such that
\begin{align}
\lim_{t\rightarrow T^-} \int_{\mathbb{R}^N}[|\nabla u(x,t)|^2+|\nabla h(|u(x,t)|^2)|^2)]dx=+\infty.
\end{align}
 }

About the topics on the global existence and blowup phenomena of the classical semilinear Schr\"{o}dinger equation, Glassey studied the following Cauchy problem
\begin{equation}
\label{2} \left\{
\begin{array}{lll}
iu_t=\Delta u+F(|u|^2)u, \quad x\in \mathbb{R}^N, \ t>0\\
u(x,0)=u_0(x),\quad x\in \mathbb{R}^N
\end{array}\right.
\end{equation}
in his famous paper \cite{Glassey}. He showed that: If there exists a constant $c_N>1+\frac{2}{N}$ such that $sF(s)\geq c_N G(s)\geq 0$ for all $s\geq 0$, where $G(s)=\int_0^s F(\eta)d\eta$, then the solution will blow up in finite time for certain initial $u_0$. If $F(|u|^2)u=\pm|u|^{\frac{4}{N-2}}u$, (\ref{2}) is in the energy critical case. We also can refer to \cite{Berestycki, Cazenave, Duyckaerts, Kenig1, Kenig, Killip, Killip1, Visan} and the references therein. However, there are very few results on the global existence and blowup phenomena of qusilinear Schr\"{o}dinger equations, we can refer to \cite{Bouard, Guo, SongWang1}.

This paper parallels to \cite{SongWang1}. Recently, in \cite{SongWang1}, we studied the following Cauchy problem
\begin{equation}
\label{1'} \left\{
\begin{array}{lll}
iu_t=\Delta u+2uh'(|u|^2)\Delta h(|u|^2)+F(|u|^2)u,\quad x\in \mathbb{R}^N, \ t>0\\
u(x,0)=u_0(x),\quad x\in \mathbb{R}^N,
\end{array}\right.
\end{equation}
and provided sufficient conditions on the blowup in finite time and global existence of the solution to (\ref{1'}) in the case of
\begin{align}
\lim_{s\rightarrow +\infty} \frac{G(s)}{s^{\frac{2^*}{2}}+[h(s)]^{2^*}}=0,\quad \lim_{s\rightarrow +\infty} \frac{F(s)s}{s^{\frac{2^*}{2}}+[h(s)]^{2^*}}=0.\label{329w1}
\end{align}
Naturally, we are interested in the following question: What's about conditions on the blowup and global existence of the solution to (\ref{1'}) in the critical case of
\begin{align}
\lim_{s\rightarrow +\infty} \frac{|G(s)|}{s^{\frac{2^*}{2}}+[h(s)]^{2^*}}=a>0\quad {\rm or}\quad \lim_{s\rightarrow +\infty} \frac{|F(s)s|}{s^{\frac{2^*}{2}}+[h(s)]^{2^*}}=b>0?
\label{329w2}
\end{align}
This is the first reason why we consider (\ref{1}) which satisfies (\ref{329w2}). Other reasons are as follows: We established pseudoconformal conservation law for the solution (\ref{1'}) in \cite{SongWang1}, which is essential for the study of the asymptotic behavior for the solution. Naturally, we hope to get pseudoconformal conservation law for the solution of (\ref{1}). It is well known that Morawetz estimate is an important tool to construct scattering operator on the energy space. A deeper question is: What is the relationship between pseudoconformal conservation law and Morawetz estimate? To solve this question, we will establish Morawetz estimate for the solution of (\ref{1}) based on pseudoconformal conservation law. Meanwhile, basing on pseudoconformal conservation law, we give some spacetime bound estimates for the global solution of (\ref{1}A), which reveals the relationship between spacetime bound and pseudoconformal conservation law. These are our ideas which generated very recently, we also can refer to our paper \cite{Song1}.

There are two main goals of this paper: One is to establish conditions on blowup and global existence of the solution to (\ref{1}), another is to give Morawetz estimates and spacetime bounds for the global solution of (\ref{1}A) based on pseudoconformal conservation law. Before we state our results, we define the mass and energy of (\ref{1}) as follows.

(i) Mass: $$ m(u)=\left(\int_{\mathbb{R}^N}|u(\cdot,t)|^2dx\right)^{\frac{1}{2}}:=[M(u)]^{\frac{1}{2}};$$

(ii) Energy : $$E(u)=\frac{1}{2}\int_{\mathbb{R}^N}[|\nabla u|^2+|\nabla h(|u|^2)|^2-G(|u|^2)\pm\frac{A}{2^*}[h(|u|^2)]^{2^*}]dx.$$

We will prove mass and energy conservation laws in Section 2.

We use $C_s$ to denote the best constant in the Sobolev's inequality
\begin{align}
\int_{\mathbb{R}^N}w^{2^*}dx\leq C_s\left(\int_{\mathbb{R}^N}|\nabla w|^2dx\right)^{\frac{2^*}{2}}\quad {\rm for \ any}\quad w\in H^1(\mathbb{R}^N).\label{zjcs}
\end{align}

The first theorem is about sufficient conditions on the global existence of the solution to (\ref{1}A) and blowup of the solution to (\ref{1}B).

{\bf Theorem 1} {\bf (A). The conditions on the global existence of the solution to (\ref{1}A)}. {\it  Let $u(x,t)$ be the solution of {\bf (\ref{1}A)} with $u_0\in X$. Here
\begin{align}
X=\{w\in H^1(\mathbb{R}^N),\quad \int_{\mathbb{R}^N}|\nabla h(|w|^2)|^2dx<+\infty\}.\label{kongjianziji}
\end{align}
Assume that (\ref{3251}) holds, $F(s)=F_1(s)-F_2(s)$, $F_1(s)\geq 0$ or $F_1(s)$ changes sign for $s\geq 0$, $F_2(s)\geq 0$ for $s\geq 0$, correspondingly, $G(s)=G_1(s)-G_2(s)$.
Then the solution of (\ref{1}A) is global existence for any initial
 $u_0\in X$ satisfying $0<M(u_0)<+\infty$ and $0<E(u_0)<+\infty$ in one of the following cases:

 {\bf Case (a).} There exists constant $\bar{m}_1\geq 0$ such that
 \begin{align}
 |G_1(s)|\leq \bar{m}_1s+G_2(s)\quad {\rm for}\quad s\geq 0;\label{330x1}
 \end{align}

{\bf Case (b).} There exists constant $\bar{m}_2\geq0$ such that
 \begin{align}
 |G_1(s)|\leq \bar{m}_2s+\frac{A}{2^*}[h(s)]^{2^*}\quad {\rm for}\quad s\geq 0.\label{330x2}
 \end{align}
}

{\bf (B). The conditions on blowup of the solution to (\ref{1}B)}.  {\it Assume that $u(x,t)$ is the solution to {\bf (\ref{1}B)} with $u_0\in X$, $xu_0\in L^2(\mathbb{R}^N)$, $\Im \int_{\mathbb{R}^N}\bar{u}_0(x\cdot \nabla u_0)dx>0$, (\ref{3251}) holds, and there exists constant $k\geq -\frac{1}{2}$ such that $sh''(s)\leq kh'(s)$. Then the solution $u(x,t)$ will blow up in finite time in one of the following cases:

{\bf Case (c).}  $-\infty<E(u_0)\leq 0$ and
\begin{align}
NF(s)s-2[(k+1)N+1]G(s)\geq 0,\quad Nh'(s)s-\frac{2[(k+1)N+1]}{2^*}h(s)\geq 0\label{416x1}
\end{align}
for $s\geq 0$.

{\bf Case (d).} There exist $\tilde{M}_1>0$ and $\tilde{M}_2\geq 0$ such that
\begin{align}
&Nh'(s)s-\frac{2[(k+1)N+1]}{2^*}h(s)\geq \tilde{M}_1h(s),\label{330x3}\\
& [|2[(k+1)N+1]||G(s)|+N|F(s)|s]\leq A\tilde{M}_1[h(s)]^{2^*}+\tilde{M}_2s\quad {\rm for}\ s\geq 0,\label{328w1}\\
&-\infty<2[(2k+1)N+2]E(u_0)+\tilde{M}_2M(u_0)\leq 0.\label{328w2}
\end{align}

}

The following pseudoconformal conservation law is essential for the study of the asymptotic behaviour of the solution of (\ref{1}A), which is inspired by \cite{Ginibre1, Ginibre2, Ginibre3, Ginibre4}.

{\bf Theorem 2.} ({\bf Pseudoconformal conservation law}.) {\it Let $u(x,t)$ be the global solution of ({\bf \ref{1}A}) in energy space $X$, $u_0\in X$ and $xu_0\in L^2(\mathbb{R}^N)$.
 Then
\begin{align}
&\int_{\mathbb{R}^N}|(x-2it\nabla)u|^2dx+4t^2\int_{\mathbb{R}^N}[|\nabla h(|u|^2)|^2-G(|u|^2)+ \frac{A}{2^*}[h|u|^2]^{2^*}]dx\nonumber\\
&=\int_{\mathbb{R}^N}|xu_0|^2dx+4\int_0^t\tau\theta(\tau)d\tau.\label{2211}
\end{align}
Here
\begin{align}
\theta(t)&=-\int_{\mathbb{R}^N}4N[2h''(|u|^2)h'(|u|^2)|u|^2+( h'(|u|^2))^2]|u|^2|\nabla u|^2dx\nonumber\\
&\quad-\int_{\mathbb{R}^N} [(N+2)G(|u|^2)-NF(|u|^2)|u|^2]dx\nonumber\\
&\quad-\int_{\mathbb{R}^N} A[h(|u|^2)]^{2^*-1}[Nh'(|u|^2)|u|^2-\frac{N+2}{2^*}h(|u|^2)]dx.\label{2212}
\end{align}
}

Basing on pseudoconformal conservation law, we establish Morawetz estimates for the solution of (\ref{1}A).

{\bf Theorem 3.} ({\bf Morawetz estimates for the solution of (\ref{1}A).}) {\it Let $u(x,t)$ be the global solution of {\bf (\ref{1}A)} with $u_0\in X$ satisfying $xu_0\in L^2(\mathbb{R}^N)$, $0<M(u_0)<+\infty$ and $0<E(u_0)<+\infty$. Assume that (\ref{3251}) holds, $F(s)=F_1(s)-F_2(s)$, $F_1(s)\geq 0$ or $F_1(s)$ changes sign for $s\geq 0$, $F_2(s)\geq 0$ for $s\geq 0$, correspondingly, $G(s)=G_1(s)-G_2(s)$.  Suppose that there exist constants $m_1, m'_1, m_2, m'_2>0$, $0<\gamma_1, \tilde{\gamma}_1<1$ and $\gamma_2,\tilde{\gamma}_2>1$ such that
\begin{align}
&\frac{2^*(1-\gamma_1)}{2(\gamma_2-\gamma_1)}=\frac{2^*(1-\tilde{\gamma}_1)}{2(\tilde{\gamma}_2-\tilde{\gamma}_1)}=1,\quad
M_r(u_0):=\sum_{j=1}^2(m_j\|u_0\|^2_{L^2})^{\frac{2}{N}}(m'_jC_s)^{\frac{N-2}{N}}<1,\label{6261}\\
& [|G_1(s)|]^{\gamma_1}\leq m_1s ,\quad [|G_1(s)|]^{\gamma_2}\leq m'_1[h(s)]^{2^*}\ {\rm for}\ 0\leq s\leq 1,\label{6262}\\
& [|G_1(s)|]^{\tilde{\gamma}_1}\leq m_2s ,\quad [|G_1(s)|]^{\tilde{\gamma}_2}\leq m'_2[h(s)]^{2^*}\ {\rm for}\  s>1.\label{6263}
\end{align}

{\bf Case (1).} Assume that $2h''(s)h'(s)s+[h'(s)]^2\geq 0$, $Nh'(s)s-\frac{N+2}{2^*}h(s)\geq 0$, $(N+2)G_1(s)-NF_1(s)s\geq 0$ and $[NF_2(s)s-(N+2)G_2(s)]\geq 0$ for all $s\geq 0$.
Then

{\bf Estimate (C):}

\begin{align}
\int_0^{+\infty}\int_{\mathbb{R}^N}\frac{\left[|\nabla h(|u|^2)|^2+|G_1(|u|^2)|+|G_2(|u|^2)|+\frac{A}{2^*}[h(|u|^2)]^{2^*}\right]^{\theta}}{n_1(x,t)}dxdt\leq M_1(u_0,\theta)\label{2211'}
\end{align}
for $\frac{1}{2}<\theta<1$ and $n_1(x,t)\geq \tilde{n}_1(x)\geq 0$, where $\frac{1}{\tilde{n}_1(x)}\in L^{\frac{1}{1-\theta}}(\mathbb{R}^N)$, ;

{\bf Estimate (D):}
\begin{align}
\int_0^{+\infty}\int_{\mathbb{R}^N}\frac{t^2\left[|\nabla h(|u|^2)|^2+|G_1(|u|^2)|+|G_2(|u|^2)|+\frac{A}{2^*}[h(|u|^2)]^{2^*}\right]}{n_2(x,t)}dxdt\leq M_2(u_0,\mu), \label{2222}
\end{align}
where $n_2(x,t)\geq \tilde{n}_2(x)+t^{\mu}\geq 0$ for any $x\in\mathbb{R}^N$ and $t\geq 0$,  and $1<\mu<3$ if $\tilde{n}_2(x)\geq 0$, while $1<\mu$ if $\tilde{n}_2(x)\geq b>0$.

Especially, let $n_2(x,t)=t^2$, then

{\bf Estimate (E): }
\begin{align}
\int_0^{+\infty}\int_{\mathbb{R}^N}\left[|\nabla h(|u|^2)|^2+|G_1(|u|^2)|+|G_2(|u|^2)|+\frac{A}{2^*}[h(|u|^2)]^{2^*}\right]dxdt\leq M_3(u_0).\label{2223}
\end{align}

{\bf Case (2).} Assume that

(i) $-k_1[h'(s)]^2\leq 2h''(s)h'(s)s+[h'(s)]^2\leq 0$ for some $k_1>0$;

(ii) $-k_2h(s)\leq Nh'(s)s-\frac{N+2}{2^*}h(s)\leq 0$ for some $k_2>0$;

(iii) $-k_3|G_1(s)|\leq (N+2)G_1(s)-NF_1(s)s\leq 0$ for some $k_3>0$;

(iv) $-k_4G_2(s)\leq NF_2(s)s-(N+2)G_2(s)\leq 0$ for some $k_4>0$.

Let
\begin{align}
l=\max(Nk_1, k_2, k_3, k_4).\label{311xj2}
\end{align}

Then

{\bf Estimate (F):}
\begin{align}
\int_0^{+\infty}\int_{\mathbb{R}^N}\frac{t^2\left[|\nabla h(|u|^2)|^2+|G_1(|u|^2)|+|G_2(|u|^2)|+\frac{A}{2^*}[h(|u|^2)]^{2^*}\right]}{n_3(x,t)}dxdt\leq M_4(u_0,\mu,l)\label{2224}
\end{align}
 Here $n_3(x,t)\geq (\tilde{n}_3(x)+t)^{\mu}\geq 0$ for all $x\in \mathbb{R}^N$ and $t\geq 0$, and $3>\mu>1+\frac{l[1+M_r(u_0)]}{1-M_r(u_0)}$ if $\tilde{n}_3(x)\geq 0$ for all $x\in \mathbb{R}^N$, while $\mu>1+\frac{l[1+M_r(u_0)]}{1-M_r(u_0)}$ if $\tilde{n}_3(x)\geq c>0$ for all $x\in \mathbb{R}^N$.

Especially, if $n_3(x,t)=t^2$ and $l<\frac{1-M_r(u_0)}{1+M_r(u_0)}$, then

{\bf Estimate (G):}
\begin{align}
\int_0^{+\infty}\int_{\mathbb{R}^N}\left[|\nabla h(|u|^2)|^2+|G_1(|u|^2)|+|G_2(|u|^2)|+\frac{A}{2^*}[h(|u|^2)]^{2^*}\right]dxdt\leq M_5(u_0,l).\label{2225}
\end{align}
}

The following spacetime bounds for the solution of (\ref{1}A) are also based on pseudoconformal conservation law.

{\bf Theorem 4.} ({\bf Spacetime bounds for the solution of (\ref{1}A).}) {\it Suppose that $h(s)$, $F(s)$ and $G(s)$ satisfy the assumptions of Theorem 3. Then

{\bf Bound (H):}
\begin{align}
&\left(\int_0^{+\infty}\left(\int_{\mathbb{R}^N}\left[|\nabla h(|u|^2)|^2+|G_1(|u|^2)|+|G_2(|u|^2)|+\frac{A}{2^*}[h(|u|^2)]^{2^*}\right]dx\right)^pdt\right)^{\frac{1}{p}}\nonumber\\
&\leq C(u_0,p).\label{37w1}
\end{align}
Here $p>\frac{1}{2}$ in Case (1), and
$$
p>\max\left(\frac{1}{2}, \frac{[1-M_r(u_0)]}{[2(1-M_r(u_0))-l(1+M_r(u_0))]}\right),\quad 0<l<\frac{2[1-M_r(u_0)]}{[1+M_r(u_0)]}
$$
in Case (2).

{\bf Bound (I):}
\begin{align}
\|G_1(|u|^2)\|_{L^q(\mathbb{R}^+)L^r(\mathbb{R}^N)}&=\left(\int_0^{+\infty}\left(\int_{\mathbb{R}^N}[|G_1(|u|^2)|]^rdx\right)^{\frac{q}{r}}dt\right)^{\frac{1}{q}}\nonumber\\
&\leq C(u_0,r,q,\gamma_1,\gamma_2,\tilde{\gamma}_1,\tilde{\gamma}_2).\label{37w2}
\end{align}
Here $1\leq r<\gamma_2$, $1\leq r<\tilde{\gamma}_2$, and
$$q>\frac{r(\gamma_2-\gamma_1)}{2^*(r-\gamma_1)},\quad q>\frac{r(\tilde{\gamma}_2-\tilde{\gamma}_1)}{2^*(r-\tilde{\gamma}_1)}$$  in Case (1),
$$q>\frac{2r(\gamma_2-\gamma_1)[1-M_r(u_0)]}{2^*(r-\gamma_1)[2(1-M_r(u_0))-l(1+M_r(u_0))]},$$
$$
q>\frac{2r(\tilde{\gamma}_2-\tilde{\gamma}_1)[1-M_r(u_0)]}{2^*(r-\tilde{\gamma}_1)[2(1-M_r(u_0))-l(1+M_r(u_0))]},
$$
$0<l<\frac{2[1-M_r(u_0)]}{[1+M_r(u_0)]}$ in Case (2).
}

{\bf Remark 1.1.} All the results on (\ref{1}A) are true in the special case of $F_1(s)\equiv 0$ or $F_2(s)\equiv 0$.

The organization of this paper is as follows. In Section 2, we will prove mass and energy conservation laws and some equalities. In Section 3, we will prove Theorem 1, obtain sufficient conditions on global existence of the solution to (\ref{1}A) and those on the blowup of the solution to (\ref{1}B). In Section 4, we establish pseudoconformal conservation law and Morawetz estimates for the solution of (\ref{1}A). In Section 5, we give spacetime bound estimates for the solution of (\ref{1}A).

\section{Preliminaries}
\qquad In the sequels, we will use $C$, $C'$, and so on, to denote different constants, the values of them may vary occurrence to occurrence.

We will prove a lemma in this section.

{\bf Lemma 2.1.} {\it Assume that $u$ is the solution to (\ref{1}). Then in the time interval $[0,t]$ when it exists, $u$ satisfies

(i) Mass conversation: $$ M(u):=\|u(\cdot,t)\|_2^2=\|u_0\|^2_2=M(u_0);$$

(ii) Energy conversation: $$E(u)=\frac{1}{2}\int_{\mathbb{R}^N}[|\nabla u|^2+|\nabla h(|u|^2)|^2-G(|u|^2)\pm \frac{A}{2^*}[h(|u|^2)]^{2^*}]dx=E(u_0);$$

(iii) $$\frac{d}{dt} \int_{\mathbb{R}^N}|x|^2|u|^2dx=-4\Im \int_{\mathbb{R}^N} \bar{u}(x\cdot \nabla u)dx;$$

(iv) \begin{align}
\frac{d}{dt} \Im \int_{\mathbb{R}^N} \bar{u}(x\cdot \nabla u)dx&=-2\int_{\mathbb{R}^N}|\nabla u|^2dx-(N+2)\int_{\mathbb{R}^N}|\nabla h(|u|^2)|^2dx\nonumber\\
&\quad -8N\int_{\mathbb{R}^N}h''(|u|^2)h'(|u|^2)|u|^4|\nabla u|^2dx\nonumber\\
&\quad+N\int_{\mathbb{R}^N}[|u|^2F(|u|^2)-G(|u|^2)]dx\nonumber\\
&\quad\mp N\int_{\mathbb{R}^N}A[h(|u|^2)]^{2^*-1}[h'(|u|^2)|u|^2-\frac{1}{2^*}h(|u|^2)]dx.\label{3261}
\end{align}

}

{\bf Proof:} (i) Multiplying (\ref{1}) by $2\bar{u}$, taking the imaginary part of the result, we get
\begin{align}
\frac{\partial }{\partial t}|u|^2=\Im(2\bar{u}\Delta u) =\nabla \cdot (2\Im \bar{u}\nabla u).\label{10121}
\end{align}
Integrating it over $\mathbb{R}^N\times [0,t]$, we obtain
$$ \int_{\mathbb{R}^N}|u|^2dx=\int_{\mathbb{R}^N}|u_0|^2dx,$$
which implies mass conservation.

(ii)  Multiplying (\ref{1}) by $2\bar{u}_t$, taking the real part of the result, then integrating it over $\mathbb{R}^N\times [0,t]$, we have
\begin{align*}
&\quad\int_{\mathbb{R}^N}[|\nabla u|^2+|\nabla h(|u|^2)|^2-G(|u|^2)\pm \frac{A}{2^*}[h(|u|^2)]^{2^*}]dx\nonumber\\
&=\int_{\mathbb{R}^N}[|\nabla u_0|^2+|\nabla h(|u_0|^2)|^2-G(|u_0|^2)\pm \frac{A}{2^*}[h(|u_0|^2)]^{2^*}]dx,
\end{align*}
which means energy conservation.

(iii) Multiplying (\ref{10121}) by $|x|^2$ and integrating it over $\mathbb{R}^N$, we obtain
\begin{align*}
\frac{d}{dt}\int_{\mathbb{R}^N}|x|^2|u|^2dx&=\int_{\mathbb{R}^N}|x|^2\nabla \cdot(2\Im (\bar{u}\nabla u))dx
=-4\Im \int_{\mathbb{R}^N}\bar{u}(x\cdot \nabla u)dx.
\end{align*}

(iv) Denote $u(x,t)=a(x,t)+ib(x,t)$, i.e.,  $a(x,t)=\Re u(x,t)$ and $b(x,t)=\Im u(x,t)$. Then
\begin{align*}
&\Im \bar{u}(x\cdot\nabla u)=\Im\left((a-ib)[(x\cdot\nabla a)+i(x\cdot\nabla b)]\right)=a(x\cdot \nabla b)-b(x\cdot\nabla a),\\
& \frac{d}{dt}\Im \bar{u}(x\cdot \nabla u)= \sum_{k=1}^N[x_k(b_t)_{x_k}a-x_k(a_t)_{x_k}b]+\sum_{k=1}^N(x_kb_{x_k}a_t-x_ka_{x_k}b_t),
\end{align*}
 and

\begin{align*}
&\quad \frac{d}{dt}\Im \int_{\mathbb{R}^N}\bar{u}(x\cdot \nabla u)dx\nonumber\\
 &=\int_{\mathbb{R}^N} \sum_{k=1}^N[x_k(b_t)_{x_k}a-x_k(a_t)_{x_k}b]dx
+\int_{\mathbb{R}^N} \sum_{k=1}^N (x_ka_{x_k}\Delta a+x_kb_{x_k}\Delta b)dx\nonumber\\
&\quad+\frac{1}{2}\int_{\mathbb{R}^N}\sum_{k=1}^N x_k(|u|^2)_{x_k}[2h'(|u|^2)\Delta h(|u|^2)+F(|u|^2)\mp A [h(|u|^2)]^{2^*-1}h'(|u|^2)]dx\nonumber\\
&=N\int_{\mathbb{R}^N}(a_tb-ab_t)dx+\int_{\mathbb{R}^N}\sum_{k=1}^N(x_kb_{x_k}a_t-x_ka_{x_k}b_t)dx+\frac{N-2}{2}\int_{\mathbb{R}^N}|\nabla u|^2dx\nonumber\\
&\quad+\frac{N-2}{2}\int_{\mathbb{R}^N}|\nabla h(|u|^2)|^2dx
-\frac{N}{2}\int_{\mathbb{R}^N}G(|u|^2)dx\pm \frac{NA}{22^*}\int_{\mathbb{R}^N}[h(|u|^2)]^{2^*}dx\nonumber\\
&=N\int_{\mathbb{R}^N}\left([a\Delta a+b\Delta b]+|u|^2[2h'(|u|^2)\Delta h(|u|^2)+F(|u|^2)\mp A[h(|u|^2)]^{2^*-1}h'(|u|^2)]\right)dx\nonumber\\\
&\quad +(N-2)\int_{\mathbb{R}^N}[|\nabla u|^2+|\nabla h(|u|^2)|^2]dx-N\int_{\mathbb{R}^N}G(|u|^2)dx\pm \frac{NA}{2^*}\int_{\mathbb{R}^N}[h(|u|^2)]^{2^*}dx\nonumber\\
&=-2\int_{\mathbb{R}^N}|\nabla u|^2dx-(N+2)\int_{\mathbb{R}^N}|\nabla h(|u|^2)|^2dx-8N\int_{\mathbb{R}^N}h'(|u|^2)h''(|u|^2)|u|^4|\nabla u|^2dx\nonumber\\
&\quad+N\int_{\mathbb{R}^N}[|u|^2F(|u|^2)-G(|u|^2)]dx\mp N\int_{\mathbb{R}^N}A[h(|u|^2)]^{2^*-1}[h'(|u|^2)|u|^2-\frac{1}{2^*}h(|u|^2)]dx.
\end{align*}
Lemma 2.1 is proved.\hfill $\Box$

\section{The proof of Theorem 1}
\qquad In this section,  we provide the sufficient conditions on the global existence of the solution to (\ref{1}A) and those on the blowup of the solution to (\ref{1}B).

{\bf The proof of Theorem 1:} {\bf (A).} The global existence of the solution to (\ref{1}A).

{\bf Case (a).} By mass and energy conservation laws, if (\ref{330x1}) holds, then
\begin{align*}
&\quad\int_{\mathbb{R}^N}[|\nabla u|^2+|\nabla h(|u|^2)|^2+G_2(|u|^2)+\frac{A}{2^*}[h(|u|^2)]^{2^*}]dx\nonumber\\
&=2E(u_0)+\int_{\mathbb{R}^N}G_1(|u|^2)dx\leq 2E(u_0)+\int_{\mathbb{R}^N}|G_1(|u|^2)|dx\nonumber\\
&\leq 2E(u_0)+\int_{\mathbb{R}^N}[\bar{m}_1|u|^2+G_2(|u|^2)]dx\nonumber\\
&=2E(u_0)+\bar{m}_1M(u_0)+\int_{\mathbb{R}^N}G_2(|u|^2)dx,
\end{align*}
which implies that
\begin{align}
\int_{\mathbb{R}^N}[|\nabla u|^2+|\nabla h(|u|^2)|^2+\frac{A}{2^*}[h(|u|^2)]^{2^*}]dx\leq 2E(u_0)+\bar{m}_1M(u_0).\label{330x3}
\end{align}

{\bf Case (b).} If (\ref{330x2}) holds, then
\begin{align*}
&\quad\int_{\mathbb{R}^N}[|\nabla u|^2+|\nabla h(|u|^2)|^2+G_2(|u|^2)+\frac{A}{2^*}[h(|u|^2)]^{2^*}]dx\nonumber\\
&=2E(u_0)+\int_{\mathbb{R}^N}G_1(|u|^2)dx\leq 2E(u_0)+\int_{\mathbb{R}^N}|G_1(|u|^2)|dx\nonumber\\
&\leq 2E(u_0)+\int_{\mathbb{R}^N}[\bar{m}_2|u|^2+\frac{A}{2^*}[h(|u|^2)]^{2^*}]]dx\nonumber\\
&=2E(u_0)+\bar{m}_2M(u_0)+\frac{A}{2^*}\int_{\mathbb{R}^N}[h(|u|^2)]^{2^*}]dx,
\end{align*}
which means that
\begin{align}
\int_{\mathbb{R}^N}[|\nabla u|^2+|\nabla h(|u|^2)|^2+G_2(|u|^2)]dx\leq 2E(u_0)+\bar{m}_2M(u_0).\label{330x3}
\end{align}

The solution of (\ref{1}A) is global existence under the assumptions of Theorem 1.

{\bf (B).} The blowup of the solution to (\ref{1}B).

Wherever $u$ exists, let
$$
y(t)=\Im \int_{\mathbb{R}^N}\bar{u}(x\cdot \nabla u)dx.
$$

{\bf Case (c).}
\begin{align}
\dot{y}(t)&=-2\int_{\mathbb{R}^N}|\nabla u|^2dx-(N+2)\int_{\mathbb{R}^N}|\nabla h(|u|^2)|^2dx-8N\int_{\mathbb{R}^N}h'(|u|^2)h''(|u|^2)|u|^4|\nabla u|^2dx\nonumber\\
&\quad+N\int_{\mathbb{R}^N}[|u|^2F(|u|^2)-G(|u|^2)]dx+N\int_{\mathbb{R}^N}A[h(|u|^2)]^{2^*-1}[h'(|u|^2)|u|^2-\frac{1}{2^*}h(|u|^2)]dx\nonumber\\
&\geq -2\int_{\mathbb{R}^N}|\nabla u|^2dx-(N+2+2kN)\int_{\mathbb{R}^N}|\nabla h(|u|^2)|^2dx\nonumber\\
&\quad +N\int_{\mathbb{R}^N}[|u|^2F(|u|^2)-G(|u|^2)]dx+N\int_{\mathbb{R}^N}A[h(|u|^2)]^{2^*-1}[h'(|u|^2)|u|^2-\frac{1}{2^*}h(|u|^2)]dx\nonumber\\
&=(2k+1)N\int_{\mathbb{R}^N}|\nabla u|^2dx-2[(2k+1)N+2]E(u)\nonumber\\
&\quad+\int_{\mathbb{R}^N}[NF(|u|^2)|u|^2-2[(k+1)N+1]G(|u|^2)]dx\nonumber\\
&\quad+A\int_{\mathbb{R}^N}h(|u|^2)]^{2^*-1}[Nh'(|u|^2)]|u|^2-\frac{2[(k+1)N+1]}{2^*}h(|u|^2)]dx\nonumber\\
&\geq (2k+1)N\int_{\mathbb{R}^N}|\nabla u|^2dx-2[(2k+1)N+2]E(u_0)\nonumber\\
&\geq (2k+1)N\int_{\mathbb{R}^N}|\nabla u|^2dx.\label{10141'}
\end{align}

{\bf Case (d).} Similar to (\ref{10141'}), we can obtain
\begin{align}
\dot{y}(t)&\geq (2k+1)N\int_{\mathbb{R}^N}|\nabla u|^2dx-2[(2k+1)N+2]E(u_0)-\tilde{M}_2M(u_0)\nonumber\\
&\geq (2k+1)N\int_{\mathbb{R}^N}|\nabla u|^2dx.\label{4161}
\end{align}

In both cases, we know that $y(t)$ is increasing whenever $u$ exists and $y(t)\geq y(0)>0$ under the conditions of $y(0)=\Im \int_{\mathbb{R}^N}\bar{u}_0(x\cdot u_0)dx>0$.

Setting
$$
J(t)=\int_{\mathbb{R}^N}|x|^2|u|^2dx,
$$
we have $J'(t)=-4y(t)<-4y(0)<0$. Then
$$0\leq J(t)=J(0)+\int_0^tJ'(\tau)d\tau<J(0)-4y(0)t,$$
which implies that the maximum existence interval of time for $u$ is finite, and $u$ will blow up before $\frac{J(0)}{4y(0)}$.\hfill $\Box$

We give a corollary of Theorem 1 as follows.

{\bf Corollary 3.1.} {\it 1. Let $u(x,t)$ be the solution of ({\bf \ref{1}A}) with $u_0\in X$. Suppose that there exist $0<\theta_1<1$, $0<\theta_2<1$, $q_1>1$ and $q_2>1$ such that
\begin{align}
&[|G_1(s)|]^{\theta_1}\leq c_1s,\quad [|G_1(s)|]^{q_1}\leq c'_1s+c''_1[h(s)]^{2^*}\quad {\rm for} \quad\ 0\leq s\leq 1, \label{1015x1'}\\
&[|G_1(s)|]^{\theta_2}\leq c_2s,\quad
[|G_1(s)|]^{q_2}\leq c'_2s+c''_2[h(s)]^{2^*}\quad {\rm for} \quad s>1 \label{1015x1}
\end{align}
for some nonnegative constants $c_1$, $c'_1$, $c''_1$, $c_2$, $c'_2$ and $c''_2$. Then the solution of (\ref{1}A) is global existence for any initial data
$u_0\in X$ satisfying $0<M(u_0)<+\infty$ and $0<E(u_0)<+\infty$.

2. Let $u(x,t)$ be the solution of ({\bf \ref{1}B}) with $u_0\in X$, (\ref{3251}) holds, and there exist $k\geq -\frac{1}{2}$ and $\tilde{M}_1>0$ such that
$$
sh''(s)\leq kh'(s), \quad Nh'(s)s-\frac{2[(k+1)N+1]}{2^*}h(s)\geq \tilde{M}_1h(s)\quad {\rm for} \quad s\geq 0.
$$
Suppose that there exist $0<\bar{\theta}_1<1$, $0<\bar{\theta}_2<1$, $\bar{q}_1>1$ and $\bar{q}_2>1$ such that
\begin{align}
&[|G(s)|]^{\bar{\theta}_1}\leq \bar{c}_1s,\quad [|G(s)|]^{\bar{q}_1}\leq \bar{c}'_1s+\bar{c}''_1[h(s)]^{2^*}\quad {\rm for} \quad\ 0\leq s\leq 1, \label{421}\\
&[|G(s)|]^{\bar{\theta}_2}\leq \bar{c}_2s,\quad
[|G(s)|]^{\bar{q}_2}\leq \bar{c}'_2s+\bar{c}''_2[h(s)]^{2^*}\quad {\rm for} \quad s>1 \label{422}
\end{align}
for some nonnegative constants $\bar{c}_1$, $\bar{c}'_1$, $\bar{c}''_1$, $\bar{c}_2$, $\bar{c}'_2$ and $\bar{c}''_2$ and there exist $0<\bar{\theta}_3<1$, $0<\bar{\theta}_4<1$, $\bar{q}_3>1$ and $\bar{q}_4>1$ such that
\begin{align}
&[|F(s)s|]^{\bar{\theta}_3}\leq \bar{c}_3s,\quad [|F(s)s|]^{\bar{q}_3}\leq \bar{c}'_3s+\bar{c}''_3[h(s)]^{2^*}\quad {\rm for} \quad\ 0\leq s\leq 1, \label{425}\\
&[|F(s)s|]^{\bar{\theta}_4}\leq \bar{c}_4s,\quad
[|F(s)s|]^{\bar{q}_4}\leq \bar{c}'_4s+\bar{c}''_4[h(s)]^{2^*}\quad {\rm for} \quad s>1 \label{422}
\end{align}
for some nonnegative constants $\bar{c}_3$, $\bar{c}'_3$, $\bar{c}''_3$, $\bar{c}_4$, $\bar{c}'_4$ and $\bar{c}''_4$. Then the solution of (\ref{1}B) will blow up in finite time for some initial data $u_0$. }

{\bf Proof:} 1. We only to show that: (\ref{1015x1'}) and (\ref{1015x1}) imply (\ref{330x2}). By Young inequality, taking $\epsilon$ small enough, we have
\begin{align}
|G_1(s)|&\leq C(\epsilon)|G_1(s)|^{\theta_1}+\epsilon|G_1(s)|^{q_1}\leq C(\epsilon)c_1s+\epsilon[c'_1s+c''_1[h(s)]^{2^*}]\nonumber\\
&\leq \bar{m}_2s+\frac{A}{2^*}[h(s)]^{2^*} \quad {\rm for}\quad 0\leq s\leq 1,\label{423}\\
|G_1(s)|&\leq C(\epsilon)|G_1(s)|^{\theta_2}+\epsilon|G_1(s)|^{q_2}\leq C'(\epsilon)c_1s+\epsilon[c'_2s+c''_2[h(s)]^{2^*}]\nonumber\\
&\leq \bar{m}_2s+\frac{A}{2^*}[h(s)]^{2^*} \quad {\rm for}\quad  s>1.\label{424}
\end{align}
Therefore, $|G_1(s)|\leq \bar{m}_2s+\frac{A}{2^*}[h(s)]^{2^*}$ for $s\geq 0$, (\ref{330x2}) is satisfied and the solution of (\ref{1}A) is global existence.

2. Similar to (\ref{423}) and (\ref{424}), we can get
$$|G(s)|\leq C(\epsilon)s+\epsilon [h(s)]^{2^*},\quad |F(s)s|\leq C(\epsilon)s+\epsilon [h(s)]^{2^*}\quad {\rm for}\ s\geq 0. $$
Taking $\epsilon$ small enough, we can get
$$[|2[(k+1)N+1]||G(s)|+N|F(s)|s]\leq \tilde{M}_2s+\tilde{M}_1 [h(s)]^{ 2^*}\quad {\rm for}\ s\geq 0. $$
(\ref{328w1}) is satisfied and the solution of (\ref{1}B) will blow up in finite time for the initial data $u_0$ satisfying
$xu_0\in L^2(\mathbb{R}^N)$, $\Im \int_{\mathbb{R}^N}\bar{u}_0(x\cdot \nabla u_0)dx>0$, and
$$
2[(2k+1)N+2]E(u_0)+\tilde{M}_2M(u_0)\leq 0.
$$
Corollary 3.1 is proved.\hfill$\Box$

We would like to give some examples to illustrate the results of Theorem 1. To see the difference between (\ref{1}A) and (\ref{1}B), we chose the same $h(s)$ and $F(s)$.

{\bf Example 3.1.} $h(s)=s^{\alpha}$, $\alpha\geq \frac{1}{2}$, $F(s)\equiv 0$ or $F(s)=\mp s^q$, $0<q<\alpha\cdot 2^*-1$.

For (\ref{1}A), since $|G(s)|=\frac{s^{q+1}}{q+1}\leq \bar{m}_1s+\frac{A}{2^*}s^{\alpha\cdot 2^*}$, the solution is global existence for initial data $u_0\in X$ satisfying $0<E(u_0)<+\infty$ and $0<M(u_0)<+\infty$.

For (\ref{1}B), if  $\alpha\geq \frac{1}{2}$, we can take $$k=\alpha-1\geq-\frac{1}{2},\quad \tilde{M}_1=\frac{(2^*-2)\alpha N-2}{2^*}>0.$$ By Young inequality, we have
$$
[|2[(k+1)N+1]||G(s)|+N|F(s)|s]=[\frac{2|\alpha N+1|}{q+1}+N]s^{q+1}\leq \tilde{M}_1 A s^{\alpha\cdot 2^*}+\tilde{M}_2s.
$$
If $$|x|u_0\in L^2(\mathbb{R}^N),\quad\Im \int_{\mathbb{R}^N}\bar{u}_0(x\cdot \nabla u_0)dx>0,\quad
2[(2\alpha-1)N+2]E(u_0)+\tilde{M}_2M(u_0)\leq 0,$$ then the solution of (\ref{1}B) will blow up in finite time.

{\bf Example 3.2.} $h(s)=a_1s^{\alpha_1}+...+a_ms^{\alpha_m}$, $\alpha_m\geq \frac{1}{2}$, $$F(s)=b_1s^{p_1}+...+b_ns^{p_n}-c_1s^{q_1}-...-c_rs^{q_r},$$ the coefficients $a_1$, ..., $a_m$, $b_1$, ..., $b_n$, $c_1$, ...., $c_r$ are positive, $0<\alpha_1<..<\alpha_m$,
$0<p_1<...<p_n$, $0<q_1<...<q_r$.

For (\ref{1}A), $G_1(s)=\frac{b_1}{p_1+1}s^{p_1+1}+...+\frac{b_n}{p_n+1}s^{p_n+1}$, if $p_n<\alpha_m\cdot 2^*-1$, then there exists $\bar{m}_2>0$ such that $G_1(s)\leq \bar{m}_2s+\frac{A}{2^*}[h(s)]^{2^*}$ for $s\geq 0$ and the solution is global existence for initial data $u_0\in X$ satisfying $0<E(u_0)<+\infty$ and $0<M(u_0)<+\infty$.

For (\ref{1}B), we can take $k=\alpha_m-1\geq -\frac{1}{2}$. If $(2^*\alpha_1-2\alpha_m)N-2>0$, we can take $\tilde{M}_1=N\alpha_1-\frac{2\alpha_mN+2}{2^*}$, then
\begin{align*}
|2[(k+1)N+1]||G(s)|+N|F(s)|s&=\sum_{j=1}^n[\frac{2|\alpha_mN+1|}{p_j+1}+N]s^{q_j+1}+\sum_{l=1}^r[\frac{2|\alpha_mN+1|}{q_l+1}+N]s^{q_l+1}\nonumber\\
&\leq \tilde{M}_1 As^{\alpha\cdot 2^*}+\tilde{M}_2s.
\end{align*}
If $\Im \int_{\mathbb{R}^N}\bar{u}_0(x\cdot \nabla u_0)dx>0$, $|x|u_0\in L^2(\mathbb{R}^N)$,
$2[(2\alpha_m-1)N+2]E(u_0)+\tilde{M}_2M(u_0)\leq 0$, then the solution of (\ref{1}B) will blow up in finite time.

{\bf Example 3.3.} Consider the following problem
\begin{equation}
\label{exam1} \left\{
\begin{array}{lll}
iu_t=\Delta u+2Kue^{K|u|^2}\Delta e^{K|u|^2}+ae^{L|u|^2}u-AKe^{K\cdot 2^*|u|^2}u,\ x\in \mathbb{R}^N, \ t>0\\
u(x,0)=u_0(x),\quad x\in \mathbb{R}^N.
\end{array}\right.
\end{equation}
(\ref{exam1}) is the special case of (\ref{1}) with $h(s)=e^{Ks}$, $F(s)=ae^{Ls}$ and $G(s)=\frac{a}{L}e^{Ls}$. If $L<K\cdot 2^*$ and $\frac{a}{L}<\frac{A}{2^*}$, then (\ref{330x2}) is satisfied and the solution of (\ref{exam1}) is global existence for initial data $u_0\in X$ satisfying $0<E(u_0)<+\infty$ and $0<M(u_0)<+\infty$.

As a byproduct of this example, we know that (\ref{1015x1'}) and (\ref{1015x1}) imply (\ref{330x2}). However, there exist functions $h(s)$ and $G(s)$ such that (\ref{330x2}) holds yet (\ref{1015x1'}) and (\ref{1015x1}) are not satisfied.

\section{The proofs of Theorem 2 and Theorem 3}
\subsection{Pseudoconformal conservation law}
\qquad {\bf Proof of Theorem 2:} Assume that $u$ is the global solution of (\ref{1}A), $u_0\in X$ and $xu_0\in L^2(\mathbb{R}^N)$. Using energy conservation law, we get
\begin{align}
P(t)&:=\int_{\mathbb{R}^N}|xu|^2dx+4t\Im \int_{\mathbb{R}^N}\bar{u}(x\cdot \nabla u)dx+4t^2\int_{\mathbb{R}^N}|\nabla u|^2dx\nonumber\\
&\quad+4t^2\int_{\mathbb{R}^N}|\nabla h(|u|^2)|^2dx-4t^2\int_{\mathbb{R}^N}G(|u|^2)dx+ \frac{4A}{2^*}t^2\int_{\mathbb{R}^N}[h|u|^2]^{2^*}dx\nonumber\\
&=\int_{\mathbb{R}^N}|xu|^2dx+4t\Im \int_{\mathbb{R}^N}\bar{u}(x\cdot \nabla u)dx+8t^2E(u_0).\label{692}
\end{align}
Recalling that
$$\frac{d}{dt} \int_{\mathbb{R}^N}|x|^2|u|^2dx=-4\Im \int_{\mathbb{R}^N} \bar{u}(x\cdot \nabla u)dx,$$
we obtain
\begin{align}
P'(t)&=\frac{d}{dt}\int_{\mathbb{R}^N}|xu|^2dx+4\Im \int_{\mathbb{R}^N}\bar{u}(x\cdot \nabla u)dx+4t\frac{d}{dt}\Im \int_{\mathbb{R}^N}\bar{u}(x\cdot \nabla u)dx+16tE(u_0)\nonumber\\
&=4t\frac{d}{dt}\Im \int_{\mathbb{R}^N}\bar{u}(x\cdot \nabla u)dx+16tE(u_0)\nonumber\\
&=4t\left\{-2\int_{\mathbb{R}^N}|\nabla u|^2dx-(N+2)\int_{\mathbb{R}^N}|\nabla h(|u|^2)|^2dx-8N\int_{\mathbb{R}^N}h''(|u|^2)h'(|u|^2)|u|^4|\nabla u|^2dx\right.\nonumber\\
&\quad\left.+N\int_{\mathbb{R}^N}[|u|^2F(|u|^2)-G(|u|^2)]dx- N\int_{\mathbb{R}^N}A[h(|u|^2)]^{2^*-1}[h'(|u|^2)|u|^2-\frac{1}{2^*}h(|u|^2)]dx\right\}\nonumber\\
&\qquad+8t\int_{\mathbb{R}^N}[|\nabla u|^2+|\nabla h(|u|^2)|^2-G(|u|^2)+ \frac{A}{2^*}[h(|u|^2)]^{2^*}]dx\nonumber\\
&=4t\int_{\mathbb{R}^N}-4N[2h''(|u|^2)h'(|u|^2)|u|^2+(h'(|u|^2))^2]|u|^2|\nabla u|^2dx\nonumber\\
&\quad- 4t\int_{\mathbb{R}^N}[(N+2)G(|u|^2)-N|u|^2F(|u|^2)]dx\nonumber\\
&\quad- 4t\int_{\mathbb{R}^N} A[h(|u|^2)]^{2^*-1}[Nh'(|u|^2)|u|^2-\frac{N+2}{2^*}h(|u|^2)]dx.\label{693}
\end{align}
Integrating (\ref{693}) from $0$ to $t$, we have
$$
P(t)=P(0)+4\int_0^t\tau\theta(\tau)d\tau=\int_{\mathbb{R}^N}|xu_0|^2dx+4\int_0^t\tau\theta(\tau)d\tau.
$$
That is,
\begin{align*}
&\int_{\mathbb{R}^N}|(x-2it\nabla)u|^2dx+4t^2\int_{\mathbb{R}^N}[|\nabla h(|u|^2)|^2-G(|u|^2)+ \frac{4A}{2^*}t^2[h|u|^2]^{2^*}]dx\nonumber\\
&=\int_{\mathbb{R}^N}|xu_0|^2dx+4\int_0^t\tau\theta(\tau)d\tau.
\end{align*}
Here
\begin{align*}
\theta(t)&=-\int_{\mathbb{R}^N}4N[2h''(|u|^2)h'(|u|^2)|u|^2+( h'(|u|^2))^2]|u|^2|\nabla u|^2dx\nonumber\\
&\quad-\int_{\mathbb{R}^N} [(N+2)G(|u|^2)-NF(|u|^2)|u|^2]dx\nonumber\\
&\quad-\int_{\mathbb{R}^N} A[h(|u|^2)]^{2^*-1}[Nh'(|u|^2)|u|^2-\frac{N+2}{2^*}h(|u|^2)]dx.
\end{align*}
Theorem 2 is proved. \hfill $\Box$

\subsection{Morawetz estimates based on pseudoconformal conservation law}
\qquad {\bf The proof of Theorem 3:} By energy conservation law, under the assumptions of (\ref{6261}), (\ref{6262}) and (\ref{6263}), using Young inequality, we get
\begin{align}
2E(u_0)&=\int_{\mathbb{R}^N}[|\nabla u|^2+|\nabla h(|u|^2)|^2-G_1(|u|^2)+G_2(|u|^2)+\frac{A}{2^*}[h(|u|^2)]^{2^*}]dx\nonumber\\
&=\int_{\mathbb{R}^N}[|\nabla u|^2+|\nabla h(|u|^2)|^2+G_2(|u|^2)+\frac{A}{2^*}[h(|u|^2)]^{2^*}]dx\nonumber\\
&\quad-\int_{\{|u|\leq 1\}}|G_1(|u|^2)|dx-\int_{\{|u|>1\}}|G_1(|u|^2)|dx\nonumber\\
&\geq \int_{\mathbb{R}^N}[|\nabla u|^2+|\nabla h(|u|^2)|^2+G_2(|u|^2)+\frac{A}{2^*}[h(|u|^2)]^{2^*}]dx\nonumber\\
&\quad-\left(\int_{\{|u|\leq 1\}}|G_1(|u|^2)|^{\gamma_1}dx \right)^{\frac{1}{\tilde{\tau}'_1}}\left(\int_{\{|u|\leq 1\}}|G_1(|u|^2)|^{\gamma_2}dx \right)^{\frac{1}{\tilde{\tau}_1}}\nonumber\\
&\quad-\left(\int_{\{|u|>1\}}|G_1(|u|^2)|^{\tilde{\gamma}_1}dx \right)^{\frac{1}{\tilde{\tau}'_2}}\left(\int_{\{|u|>1\}}|G_1(|u|^2)|^{\tilde{\gamma}_2}dx \right)^{\frac{1}{\tilde{\tau}_2}}\nonumber\\
&\geq \int_{\mathbb{R}^N}[|\nabla u|^2+|\nabla h(|u|^2)|^2+G_2(|u|^2)+\frac{A}{2^*}[h(|u|^2)]^{2^*}]dx\nonumber\\
&\quad-\left(\int_{\{|u|\leq 1\}}m_1|u|^2)dx \right)^{\frac{1}{\tilde{\tau}'_1}}\left(\int_{\{|u|\leq 1\}}m'_1[h(|u|^2)]^{2^*}dx \right)^{\frac{1}{\tilde{\tau}_1}}\nonumber\\
&\quad-\left(\int_{\{|u|> 1\}}m_2|u|^2)dx \right)^{\frac{1}{\tilde{\tau}'_2}}\left(\int_{\{|u|>1\}}m'_2[h(|u|^2)]^{2^*}dx \right)^{\frac{1}{\tilde{\tau}_2}}\nonumber\\
&\geq \int_{\mathbb{R}^N}[|\nabla u|^2+|\nabla h(|u|^2)|^2+G_2(|u|^2)+\frac{A}{2^*}[h(|u|^2)]^{2^*}]dx\nonumber\\
&\quad-\sum_{j=1}^2(m_j\int_{\mathbb{R}^N}|u|^2dx)^{\frac{1}{\tilde{\tau}'_j}}\left(m'_j\int_{\mathbb{R}^N}[h(|u|^2)]^{2^*}dx \right)^{\frac{1}{\tilde{\tau}_j}}\nonumber\\
&\geq \int_{\mathbb{R}^N}[|\nabla u|^2+|\nabla h(|u|^2)|^2+G_2(|u|^2)+\frac{A}{2^*}[h(|u|^2)]^{2^*}]dx\nonumber\\
&\quad-\sum_{j=1}^2(m_j\|u_0\|^2_{L^2})^{\frac{1}{\tilde{\tau}'_j}}(m'_jC_s)^{\frac{1}{\tilde{\tau}_j}}\left(\int_{\mathbb{R}^N}|\nabla h(|u|^2)|^2dx \right)^{\frac{2^*}{2\tilde{\tau}_j}}\nonumber\\
&=\int_{\mathbb{R}^N}[|\nabla u|^2+|\nabla h(|u|^2)|^2+G_2(|u|^2)+\frac{A}{2^*}[h(|u|^2)]^{2^*}]dx\nonumber\\
&\quad-\sum_{j=1}^2(m_j\|u_0\|^2_{L^2})^{\frac{1}{\tilde{\tau}'_j}}(m'_jC_s)^{\frac{1}{\tilde{\tau}_j}}\int_{\mathbb{R}^N}|\nabla h(|u|^2)|^2dx\nonumber\\
&\geq [1-M_r(u_0)]\int_{\mathbb{R}^N}[|\nabla u|^2+|\nabla h(|u|^2)|^2+G_2(|u|^2)+\frac{A}{2^*}[h(|u|^2)]^{2^*}]dx\nonumber\\
&\geq [1-M_r(u_0)]\int_{\mathbb{R}^N}[|\nabla h(|u|^2)|^2+G_2(|u|^2)+\frac{A}{2^*}[h(|u|^2)]^{2^*}]dx.\label{326x1}
\end{align}
Here
\begin{align}
\frac{1}{\tilde{\tau}_1}&=\frac{1-\gamma_1}{\gamma_2-\gamma_1},\quad \frac{1}{\tilde{\tau}'_1}=\frac{\gamma_2-1}{\gamma_2-\gamma_1},\qquad
\frac{1}{\tilde{\tau}_2}=\frac{1-\tilde{\gamma}_1}{\tilde{\gamma}_2-\tilde{\gamma}_1},\quad
\frac{1}{\tilde{\tau}'_2}=\frac{\tilde{\gamma}_2-1}{\tilde{\gamma}_2-\tilde{\gamma}_1}.\label{415x1}
\end{align}
By the way, we obtain
\begin{align}
\int_{\mathbb{R}^N}|G_1(|u|^2)|dx&\leq \sum_{j=1}^2(m_j\|u_0\|^2_{L^2})^{\frac{1}{\tilde{\tau}'_j}}(m'_jC_s)^{\frac{1}{\tilde{\tau}_j}}\int_{\mathbb{R}^N}|\nabla h(|u|^2)|^2dx\nonumber\\
&:=M_r(u_0)\int_{\mathbb{R}^N}|\nabla h(|u|^2)|^2dx\label{326x2}
\end{align}
in the process of (\ref{326x1}).

Denoting
\begin{align}
\int_{\mathbb{R}^N}[|\nabla h(|u|^2)|^2+|G_1(|u|^2)|+G_2(|u|^2)+\frac{A}{2^*}[h(|u|^2)]^{2^*}]dx:=\int_{\mathbb{R}^N}\Psi(u)dx,\label{326x3}
\end{align}
by (\ref{326x1}) and (\ref{326x2}), we have
\begin{align}
\int_{\mathbb{R}^N}\Psi(u)dx\leq \frac{2E(u_0)[1+M_r(u_0)]}{[1-M_r(u_0)]} \ {\rm for \ any} \ t\geq 0({\rm especially \ for}\ 0\leq t\leq 1).\label{326x3}
\end{align}

To establish Morawetz estimates, the key technique is to obtain the bound for
$$
\int_{\mathbb{R}^N}[|\nabla h(|u|^2)|^2+|G_1(|u|^2)|+G_2(|u|^2)+\frac{A}{2^*}[h(|u|^2)]^{2^*}]dx
$$
for $t\geq 1$ by using pseudoconformal conservation law.

Under the assumptions of Theorem 3, (\ref{2211}) and (\ref{2212}) become
\begin{align}
&\int_{\mathbb{R}^N}|(x-2it\nabla)u|^2dx+4t^2\int_{\mathbb{R}^N}|\nabla h(|u|^2)|^2dx-4t^2\int_{\mathbb{R}^N}G_1(|u|^2)dx+4t^2\int_{\mathbb{R}^N}G_2(|u|^2)dx\nonumber\\
&+ \frac{4A}{2^*}t^2\int_{\mathbb{R}^N}[h|u|^2]^{2^*}dx=\int_{\mathbb{R}^N}|xu_0|^2dx+4\int_0^t\tau\theta(\tau)d\tau\label{326x4}
\end{align}
and
\begin{align}
\theta(t)&=-\int_{\mathbb{R}^N}4N[2h''(|u|^2)h'(|u|^2)|u|^2+( h'(|u|^2))^2]|u|^2|\nabla u|^2dx\nonumber\\
&\quad-\int_{\mathbb{R}^N} [(N+2)G_1(|u|^2)-NF_1(|u|^2)|u|^2]dx\nonumber\\
&\quad-\int_{\mathbb{R}^N} [NF_2(|u|^2)|u|^2-(N+2)G_2(|u|^2)]dx\nonumber\\
&\quad-\int_{\mathbb{R}^N}A[h(|u|^2)]^{2^*-1}[Nh'(|u|^2)|u|^2-\frac{N+2}{2^*}h(|u|^2)]dx.\label{326x5}
\end{align}

We will discuss it in two cases.

{\bf Case (1).} $2h''(|u|^2)h'(|u|^2)|u|^2+[h'(|u|^2)]^2\geq 0$, $Nh'(|u|^2)|u|^2-\frac{N+2}{2^*}h(|u|^2)\geq 0$, $(N+2)G_1(|u|^2)-NF_1(|u|^2)|u|^2\geq 0$ and $NF_2(|u|^2)|u|^2-(N+2)G_2(|u|^2)\geq 0$.

Using (\ref{326x2}), (\ref{326x4}) and (\ref{326x5}), we obtain
\begin{align}
&\quad[1-M_r(u_0)] \left(4t^2\int_{\mathbb{R}^N}|\nabla h(|u|^2)|^2dx+4t^2\int_{\mathbb{R}^N}G_2(|u|^2)dx+\frac{4A}{2^*}t^2\int_{\mathbb{R}^N}[h(|u|^2)]^{2^*}dx\right)\nonumber\\
&\leq [1-M_r(u_0)] 4t^2\int_{\mathbb{R}^N}|\nabla h(|u|^2)|^2dx+4t^2\int_{\mathbb{R}^N}G_2(|u|^2)dx+\frac{4A}{2^*}t^2\int_{\mathbb{R}^N}[h(|u|^2)]^{2^*}dx\nonumber\\
&\leq 4t^2\int_{\mathbb{R}^N}|\nabla h(|u|^2)|^2dx-4t^2\int_{\mathbb{R}^N}G_1(|u|^2)dx+4t^2\int_{\mathbb{R}^N}G_2(|u|^2)dx+ \frac{4A}{2^*}t^2\int_{\mathbb{R}^N}[h|u|^2]^{2^*}dx\nonumber\\
&\leq \int_{\mathbb{R}^N}|xu_0|^2dx:=C(u_0),\label{326w1}
\end{align}
which means that
\begin{align}
\int_{\mathbb{R}^N}\Psi(u)dx\leq \frac{C(u_0)[1+M_r(u_0)]}{4[1-M_r(u_0)]t^2} \quad {\rm for \ any} \quad t\geq 1.\label{326x6}
\end{align}

In this case, Morawetz estimates can be proved below.

{\bf Estimate (C):}

Using (\ref{326x3}) and (\ref{326x6}), we get
\begin{align}
\int_0^{\infty}\int_{\mathbb{R}^N}\frac{\left[\Psi(u)\right]^{\theta}}{n_1(x,t)}dxdt&=\int_0^1\int_{\mathbb{R}^N}\frac{\left[\Psi(u)\right]^{\theta}}{n_1(x,t)}dxdt
+\int_1^{\infty}\int_{\mathbb{R}^N}\frac{\left[\Psi(u)\right]^{\theta}}{n_1(x,t)}dxdt\nonumber\\
&\leq\int_0^1\left(\int_{\mathbb{R}^N}\Psi(u)dx\right)^{\theta}\left(\int_{\mathbb{R}^N}\frac{1}{[n_1(x,t)]^{\frac{1}{1-\theta}}}dx\right)^{1-\theta}dt\nonumber\\
&\quad +\int_1^{\infty}\left(\int_{\mathbb{R}^N}\Psi(u)dx\right)^{\theta}\left(\int_{\mathbb{R}^N}\frac{1}{[n_1(x,t)]^{\frac{1}{1-\theta}}}dx\right)^{1-\theta}dt\nonumber\\
&\leq \left[\int_0^1 Cdt+\int_1^{\infty}\frac{C'}{t^{2\theta}}dt\right]\left(\int_{\mathbb{R}^N}\frac{1}{[\tilde{n}_1(x)]^{\frac{1}{1-\theta}}}dx\right)^{1-\theta}\nonumber\\
&\leq M_1(u_0,\theta),\label{222w1}
\end{align}
where
\begin{equation}
M_1(u_0,\theta)=
\left(\frac{1+M_r(u_0)}{1-M_r(u_0)}\right)^{\theta}\left([2E(u_0)]^{\theta}
+\frac{1}{2\theta-1}[\frac{C(u_0)}{4}]^{\theta}\right) \left(\int_{\mathbb{R}^N}\frac{1}{[\tilde{n}_1(x)]^{\frac{1}{1-\theta}}}dx\right)^{1-\theta}.
\end{equation}

{\bf Estimate (D):}

(a). $1<\mu<3$ if $\tilde{n}_2(x)\geq 0$, we obtain

\begin{align}
\int_0^{\infty}\int_{\mathbb{R}^N}\frac{t^2\Psi(u)}{n_2(x,t)}dxdt&=\int_0^1\int_{\mathbb{R}^N}\frac{t^2\Psi(u)}{n_2(x,t)}dxdt
+\int_1^{\infty}\int_{\mathbb{R}^N}\frac{t^2\Psi(u)}{n_2(x,t)}dxdt\nonumber\\
&\leq \int_0^1t^{2-\mu}\int_{\mathbb{R}^N}\Psi(u)dxdt+\int_1^{\infty}\frac{1}{t^{\mu}}\int_{\mathbb{R}^N}t^2\Psi(u)dxdt\nonumber\\
&\leq \int_0^1 Ct^{2-\mu}dt+\int_1^{\infty}\frac{C'}{t^{\mu}}dt\nonumber\\
&\leq \frac{1+M_r(u_0)}{1-M_r(u_0)}[\frac{2E(u_0)}{3-\mu}+\frac{C(u_0)}{4(\mu-1)}].
\end{align}

(b). $1<\mu$ if $\tilde{n}_2(x)\geq b>0$, we get
\begin{align}
&\quad\int_0^{\infty}\int_{\mathbb{R}^N}\frac{t^2\Psi(u)}{n_2(x,t)}dxdt\leq \int_0^1\frac{Ct^2}{b}dt+\int_1^{\infty}\frac{C'}{t^{\mu}}dt\nonumber\\
&\leq
\frac{1+M_r(u_0)}{1-M_r(u_0)}[\frac{2E(u_0)}{3b}+\frac{C(u_0)}{4(\mu-1)}].
\end{align}

 Especially, if $n_2(x,t)\equiv t^2$, we have

 {\bf Estimate (E):}
\begin{align}
&\quad\int_0^{\infty}\int_{\mathbb{R}^N}\left[|\nabla h(|u|^2)|^2+|G_1(|u|^2)|+G_2(|u|^2)+\frac{A}{2^*}[h(|u|^2)]^{2^*}\right]dxdt\nonumber\\
&\leq
\frac{1+M_r(u_0)}{1-M_r(u_0)}[2E(u_0)+\frac{C(u_0)}{4}].
\end{align}

{\bf Case (2).}

(i) $-k_1[h'(s)]^2\leq 2h''(s)h'(s)s+[h'(s)]^2\leq 0$ for some $k_1>0$;

(ii) $-k_2h(s)\leq Nh'(s)s-\frac{N+2}{2^*}h(s)\leq 0$ for some $k_2>0$;

(iii) $ -k_3|G_1(s)|\leq (N+2)G_1(s)-NF_1(s)s\leq 0$ for some $k_3>0$;

(iv) $-k_4G_2(s)\leq NF_2(s)s-(N+2)G_2(s)\leq 0$ for some $k_4>0$.

 Recall that (\ref{326x3}), i.e.,
\begin{align*}
\int_{\mathbb{R}^N}\Psi(u)dx\leq \frac{2E(u_0)[1+M_r(u_0)]}{[1-M_r(u_0)]}
\end{align*}
for any $t\geq 0$(especially for $0<t\leq 1$).

Similar to (\ref{326w1}), using (\ref{326x2}), (\ref{326x4}) and (\ref{326x5}), we obtain
\begin{align}
&\quad[1-M_r(u_0)] \left(4t^2\int_{\mathbb{R}^N}|\nabla h(|u|^2)|^2dx+4t^2\int_{\mathbb{R}^N}G_2(|u|^2)dx+\frac{4A}{2^*}t^2\int_{\mathbb{R}^N}[h(|u|^2)]^{2^*}dx\right)\nonumber\\
&\leq C(u_0)+4l\int_0^t\tau\int_{\mathbb{R}^N}[|\nabla h(|u|^2)|^2+|G_1(|u|^2)|+G_2(|u|^2)+\frac{A}{2^*}[h(|u|^2)]^{2^*}]dxd\tau\nonumber\\
&\leq C(u_0)+4l[1+M_r(u_0)]\int_0^t\tau\int_{\mathbb{R}^N}[|\nabla h(|u|^2)|^2+G_2(|u|^2)+\frac{A}{2^*}[h(|u|^2)]^{2^*}]dxd\tau.\label{326w2}
\end{align}
Letting
\begin{align*}
B(t)=4\int_0^t\tau\int_{\mathbb{R}^N}[|\nabla h(|u|^2)|^2+G_2(|u|^2)+\frac{A}{2^*}[h(|u|^2)]^{2^*}]dxd\tau,
\end{align*}
(\ref{326w2}) implies
\begin{align}
B'(t)\leq \frac{C(u_0)}{[1-M_r(u_0)]t}+\frac{l[1+M_r(u_0)]}{[1-M_r(u_0)]t}B(t).\label{326w3}
\end{align}
Using (\ref{326x3}), applying Gronwall inequality to (\ref{326w3}), we obtain
\begin{align*}
B(t)\leq [\frac{4lE(u_0)[1+M_r(u_0)]^2+C(u_0)[1-M_r(u_0)]}{l[1-M^2_r(u_0)]}]t^{\frac{l[1+M_r(u_0)]}{1-M_r(u_0)}},
\end{align*}
and
\begin{align}
&\quad\int_{\mathbb{R}^N}[|\nabla h(|u|^2)|^2+G_2(|u|^2)+\frac{A}{2^*}[h(|u|^2)]^{2^*}]dx\nonumber\\
&\leq \frac{C(u_0)}{4[1-M_r(u_0)]t^2}+\frac{4lE(u_0)[1+M_r(u_0)]^2+C(u_0)[1-M_r(u_0)]}{4[1-M_r(u_0)]^2t^{2-\frac{l[1+M_r(u_0)]}{1-M_r(u_0)}}}\label{36xj3}
\end{align}
for $t\geq 1$. Consequently,
\begin{align}
\int_{\mathbb{R}^N}\Psi(u)dx\leq \frac{[1+M_r(u_0)]}{4[1-M_r(u_0)]}\left(\frac{C(u_0)}{t^2}+\frac{4lE(u_0)[1+M_r(u_0)]^2
+C(u_0)[1-M_r(u_0)]}{[1-M_r(u_0)]t^{2-\frac{l[1+M_r(u_0)]}{1-M_r(u_0)}}}\right) \label{327x1}
\end{align}
for any $t\geq 1$.

{\bf Estimate (F):}

(a). $3>\mu>1+\frac{l[1+M_r(u_0)]}{1-M_r(u_0)}$ if $\tilde{n}_3(x)\geq 0$, we get
\begin{align}
&\quad \int_0^{\infty}\int_{\mathbb{R}^N}\frac{t^2\Psi(u)}{n_3(x,t)}dxdt\leq \frac{2E(u_0)[1+M_r(u_0)]}{[1-M_r(u_0)]}\int_0^1t^{2-\mu}dt\nonumber\\
&\quad+\int_1^{\infty}\frac{C(u_0)[1+M_r(u_0)]}{4[1-M_r(u_0)]t^{\mu}}+
\frac{4lE(u_0)[1+M_r(u_0)]^3+C(u_0)[1-M^2_r(u_0)]}{4[1-M_r(u_0)]^2}\frac{1}{t^{\mu-\frac{l[1+M_r(u_0)]}{1-M_r(u_0)}}}dt\nonumber\\
&=\frac{[1+M_r(u_0)]}{[1-M_r(u_0)]}\left(\frac{2E(u_0)}{(3-\mu)}+\frac{C(u_0)}{4(\mu-1)}
+\frac{4lE(u_0)[1+M_r(u_0)]^2+C(u_0)[1-M_r(u_0)]}{4\{(\mu-1)[1-M_r(u_0)]-l[1+M_r(u_0)]\}}\right).\label{32x4}
\end{align}

(b). $\mu>1+\frac{l[1+M_r(u_0)]}{1-M_r(u_0)}$ if $\tilde{n}_3(x)\geq c>0$. Similar to (\ref{32x4}), we get
\begin{align}
 \int_0^{\infty}\int_{\mathbb{R}^N}\frac{t^2\Psi(u)}{n_3(x,t)}dxdt&\leq \frac{[1+M_r(u_0)]}{[1-M_r(u_0)]}\left(\frac{2E(u_0)}{3c^{\mu}}+\frac{C(u_0)}{4(\mu-1)}\right.\nonumber\\
&\qquad\quad\left.+\frac{4lE(u_0)[1+M_r(u_0)]^2+C(u_0)[1-M_r(u_0)]}{4\{(\mu-1)[1-M_r(u_0)]-l[1+M_r(u_0)]\}}\right)\label{32xxj2}
\end{align}

{\bf Estimate (G):}

 Especially, if $n_3(x,t)\equiv t^2$, $l<\frac{1-M_r(u_0)}{1+M_r(u_0)}$, by the discussions above, we have
\begin{align*}
&\quad\int_0^{\infty}\int_{\mathbb{R}^N}\Psi(u)dxdt\nonumber\\
&\leq
\frac{[1+M_r(u_0)]}{[1-M_r(u_0)]}\left(2E(u_0)+\frac{C(u_0)}{4}+\frac{4lE(u_0)[1+M_r(u_0)]^2+C(u_0)[1-M_r(u_0)]}{4\{[1-M_r(u_0)]-l[1+M_r(u_0)]\}}\right).
\end{align*}

{\bf Remark 4.1.} The assumptions of Case (2) can be weaken as: Assume that at least one of (i)--(iv) holds. For example, we can take $l=Nk_1$ if (i) holds, while $Nh'(s)s-\frac{N+2}{2^*}h(s)\geq 0$, $(N+2)G_1(s)-NF_1(s)s\geq 0$ and $NF_2(s)s-(N+2)G_2(s)\geq 0$; we can take $l=\max(Nk_1,k_2)$ if (i) and (ii) hold, while $NF_1(s)s-(N+2)G_1(s)\leq 0$ and $NF_2(s)s-(N+2)G_2(s)\geq 0$, and so on.

By (\ref{326x6}) and (\ref{327x1}), mass and energy conservation laws we can get the decay rate and asymptotic behavior for the solution as $t\rightarrow +\infty$, which can be states as follows.

{\bf Proposition 4.1.} {\it Assume that $u$ is the global solution of ({\bf \ref{1}A}) and the assumptions of Theorem 2 hold. Then  as $t\rightarrow +\infty$, the decay rate of $u$ satisfies
$$
\int_{\mathbb{R}^N}[|\nabla h(|u|^2)|^2+|G_1(|u|^2)|+G_2(|u|^2)+\frac{A}{2^*}[h(|u|^2)]^{2^*}]dx\leq \frac{C}{t^2}
$$
in Case (1) and
$$
\int_{\mathbb{R}^N}[|\nabla h(|u|^2)|^2+|G_1(|u|^2)|+G_2(|u|^2)+\frac{A}{2^*}[h(|u|^2)]^{2^*}]dx\leq \frac{C}{t^{2-\frac{l[1+M_r(u_0)]}{1-M_r(u_0)}}}
$$
in Case (2). Consequently,
\begin{align}
\lim_{t\rightarrow+\infty}\int_{\mathbb{R}^N}|\nabla u|^2dx=2E(u_0),\quad \lim_{t\rightarrow+\infty}\int_{\mathbb{R}^N}[|u|^2+|\nabla u|^2]dx=M(u_0)+2E(u_0).
\end{align}
}

We would like to give two examples to illustrate the results of Theorem 3.

{\bf Example 4.1.}  $h(s)=s^{\alpha}$, $\alpha\geq \frac{1}{2}$, $F(s)=s^{2\alpha-1+\frac{2}{N}}-s^q$,
$G(s)=\frac{s^{2\alpha+\frac{2}{N}}}{2\alpha+\frac{2}{N}}-\frac{s^{q+1}}{q+1}$,
$$
\gamma_1=\frac{1}{2\alpha+\frac{2}{N}},\quad \gamma_2=\frac{\alpha\cdot 2^*}{2\alpha+\frac{2}{N}},\quad \frac{2^*(1-\gamma_1)}{2(\gamma_2-\gamma_1)}=1.
$$
Under certain assumptions, if $q\geq \frac{2}{N}$, then $NF_2(s)s-(N+2)G_2(s)=(N-\frac{N+2}{q+1})s^{q+1}\geq 0$,
$$
\int_{\mathbb{R}^N}[|\nabla (|u|^{2\alpha})|^2+|u|^{4\alpha+\frac{4}{N}}+|u|^{2q+2})+\frac{A}{2^*}|u|^{2\alpha\cdot 2^*}]dx\leq \frac{C}{t^2};
$$
If $q<\frac{2}{N}$, then $NF_2(s)s-(N+2)G_2(s)=(N-\frac{N+2}{q+1})s^{q+1}<0$,
$$
\int_{\mathbb{R}^N}[|\nabla (|u|^{2\alpha})|^2+|u|^{4\alpha+\frac{4}{N}}+|u|^{2q+2})+\frac{A}{2^*}|u|^{2\alpha\cdot2^*}]dx\leq \frac{C}{t^{2-\frac{l[1+M_r(u_0)]}{1-M_r(u_0)}}}.
$$
And
$$
\int_0^{+\infty}\int_{\mathbb{R}^N}[|\nabla (|u|^{2\alpha})|^2+|u|^{4\alpha+\frac{4}{N}}+|u|^{2q+2})+\frac{A}{2^*}|u|^{2\alpha\cdot2^*}]dxdt\leq C.
$$

{\bf Example 4.2.} $h(s)=s^{\alpha_1}+s^{\alpha_2}$, $\frac{1}{2}\leq \alpha_1<\alpha_2$,
\begin{align*}
& F(s)=s^{2\alpha_1-1+\frac{2}{N}}+s^{p_1}+...+s^{p_m}+s^{2\alpha_2-1+\frac{2}{N}},\\
& G(s)=\frac{s^{2\alpha_1+\frac{2}{N}}}{2\alpha_1+\frac{2}{N}}+\frac{s^{p_1+1}}{p_1+1}
+...+\frac{s^{p_m+1}}{p_m+1}+\frac{s^{2\alpha_1+\frac{2}{N}}}{2\alpha_1+\frac{2}{N}},
\end{align*}
$2\alpha_1-1+\frac{2}{N}<p_1<...<p_m<2\alpha_2-1+\frac{2}{N}$. Obviously, $h(s)\geq s^{\alpha_1}$,
$G(s)\leq \frac{(m+2)s^{2\alpha_1+\frac{2}{N}}}{2\alpha_1+\frac{2}{N}}$ if $0\leq s\leq 1$, $h(s)\geq s^{\alpha_2}$,
$G(s)\leq \frac{(m+2)s^{2\alpha_2+\frac{2}{N}}}{2\alpha_2+\frac{2}{N}}$ if $s>1$. Since $\gamma_1=\frac{1}{2\alpha_1+\frac{2}{N}}$, $\gamma_2=\frac{\alpha_1\cdot 2^*}{2\alpha_1+\frac{2}{N}}$, $\tilde{\gamma}_1=\frac{1}{2\alpha_2+\frac{2}{N}}$ and
$\tilde{\gamma}_2=\frac{\alpha_2\cdot 2^*}{2\alpha_2+\frac{2}{N}}$, (\ref{6262}) and (\ref{6263}) hold. If the initial $u_0$ satisfies (\ref{6261}), then
$$
\int_{\mathbb{R}^N}[|\nabla (|u|^{2\alpha_1}+|u|^{2\alpha_2})|^2+G(|u|^2)+\frac{A}{2^*}[|u|^{2\alpha_1}+|u|^{2\alpha_2}]^{2^*}]dx\leq \frac{C}{t^{2-\frac{l[1+M_r(u_0)]}{1-M_r(u_0)}}}
$$
and
$$
\int_0^{+\infty}\int_{\mathbb{R}^N}[|\nabla (|u|^{2\alpha_1}+|u|^{2\alpha_2})|^2+G(|u|^2)+\frac{A}{2^*}[|u|^{2\alpha_1}+|u|^{2\alpha_2}]^{2^*}]dxdt\leq C
$$
under certain assumptions.

\section{Spacetime bound estimates based on pseudoconformal conservation law}
\qquad In this section, we give the proof of Theorem 4.

{\bf Proof of Theorem 4:}

{\bf Bound (H):} We prove (\ref{37w1}) in two cases.

{\bf Case (1).}  Recalling (\ref{326x3}) and (\ref{326x6}),
$$\int_{\mathbb{R}^N}\Psi(u)dx\leq \frac{2E(u_0)[1+M_r(u_0)]}{[1-M_r(u_0)]}\quad {\rm for}\ 0\leq t\leq 1$$
and
$$\int_{\mathbb{R}^N}\Psi(u)dx\leq \frac{C(u_0)[1+M_r(u_0)]}{4[1-M_r(u_0)]t^2}\quad {\rm for} \ t\geq1,$$
we get
\begin{align}
&\quad\left(\int_0^{+\infty}\left(\int_{\mathbb{R}^N}\Psi(u)dx\right)^pdt\right)^{\frac{1}{p}}\nonumber\\
&=\left(\int_0^1\left(\int_{\mathbb{R}^N}\Psi(u)dx\right)^pdt+\int_1^{+\infty}\left(\int_{\mathbb{R}^N}\Psi(u)dx\right)^pdt\right)^{\frac{1}{p}}\nonumber\\
&\leq \left(\int_0^1\left(\frac{2E(u_0)[1+M_r(u_0)]}{[1-M_r(u_0)]}\right)^pdt+\int_1^{+\infty}\left(\frac{C(u_0)[1+M_r(u_0)]}{4[1-M_r(u_0)]t^2}\right)^pdt\right)^{\frac{1}{p}}\nonumber\\
&\leq \tilde{c}_1\left(\int_0^1\left(\frac{2E(u_0)[1+M_r(u_0)]}{[1-M_r(u_0)]}\right)^pdt\right)^{\frac{1}{p}}
+\tilde{c}_1\left(\int_1^{+\infty}\left(\frac{C(u_0)[1+M_r(u_0)]}{4[1-M_r(u_0)]t^2}\right)^pdt\right)^{\frac{1}{p}}\nonumber\\
&\leq \frac{\tilde{c}_1[1+M_r(u_0)]}{[1-M_r(u_0)]}\left(2E(u_0)+\frac{C(u_0)}{4(2p-1)^{\frac{1}{p}}}\right).\label{38x2}
\end{align}
Here $\tilde{c}_1=1$ if $p>1$, $\tilde{c}_1=2^{\frac{1-p}{p}}$ if $\frac{1}{2}<p\leq 1$,

{\bf Case (2).} Recalling (\ref{326x3}) and (\ref{327x1}),
$$\int_{\mathbb{R}^N}\Psi(u)dx\leq \frac{2E(u_0)[1+M_r(u_0)]}{1-M_r(u_0)}$$
for $0\leq t\leq 1$ and
$$
\int_{\mathbb{R}^N}\Psi(u)dx\leq \frac{[1+M_r(u_0)]}{4[1-M_r(u_0)]}\left(\frac{C(u_0)}{t^2}+\frac{4lE(u_0)[1+M_r(u_0)]^2
+C(u_0)[1-M_r(u_0)]}{[1-M_r(u_0)]t^{2-\frac{l[1+M_r(u_0)]}{1-M_r(u_0)}}}\right)
$$
for $t\geq 1$, we obtain
\begin{align}
&\quad \left(\int_0^{+\infty}\left(\int_{\mathbb{R}^N}\Psi(u)dx\right)^pdt\right)^{\frac{1}{p}}\nonumber\\
&\leq \tilde{c}_1\left(\int_0^1\left(\frac{2E(u_0)[1+M_r(u_0)]}{1-M_r(u_0)}\right)^pdt\right)^{\frac{1}{p}}
+\tilde{c}_1\left(\int_1^{+\infty}\left(\frac{C_1}{t^2}
+\frac{C_2}{t^{2-\frac{l[1+M_r(u_0)]}{1-M_r(u_0)}}}\right)^pdt\right)^{\frac{1}{p}}\nonumber\\
&\leq \frac{2E(u_0)[1+M_r(u_0)]\tilde{c}_1}{1-M_r(u_0)}+\frac{\tilde{c}^2_1\tilde{c}_2C_1}{(2p-1)^{\frac{1}{p}}}
+\tilde{c}^2_1\tilde{c}_2C_2C_3^{\frac{1}{p}}.\label{38w1}
\end{align}
Here $\tilde{c}_2=1$ if $p<1$, $\tilde{c}_2=2^{\frac{p-1}{p}}$ if $p\geq 1$, and
 \begin{align*}
&C_1=\frac{C(u_0)[1+M_r(u_0)]}{4[1-M_r(u_0)]},\\
&C_2=\frac{4lE(u_0)[1+M_r(u_0)]^3+C(u_0)[1-M^2_r(u_0)]}{4[1-M_r(u_0)]^2},\\
&C_3=\frac{[1-M_r(u_0)]}{(2[1-M_r(u_0)]-l[1+M_r(u_0)])p-[1-M_r(u_0)]}.
\end{align*}

{\bf Bound (I):} Note that for $1\leq r<\gamma_2$, $1\leq r<\tilde{\gamma}_2$,
\begin{align}
\int_{\mathbb{R}^N}|G_1(|u|^2)|^rdx&=\int_{\{|u|\leq 1\}}|G_1(|u|^2)|^rdx+\int_{\{|u|>1\}}|G_1(|u|^2)|^rdx\nonumber\\
&\leq\left( \int_{\{|u|\leq 1\}}|G_1(|u|^2)|^{\gamma_1}dx\right)^{\frac{1}{\tau_3}}\left(\int_{\{|u|\leq 1\}}|G(|u|^2)|^{\gamma_2}dx\right)^{\frac{1}{\tau_4}}\nonumber\\
&\quad +\left(\int_{\{|u|>1\}}|G_1(|u|^2)|^{\tilde{\gamma}_1}dx\right)^{\frac{1}{\tilde{\tau}_3}}\left( \int_{\{|u|>1\}}|G(|u|^2)|^{\tilde{\gamma}_2}dx\right)^{\frac{1}{\tilde{\tau}_4}}\nonumber\\
&\leq \left( m_3\int_{\{|u|\leq 1\}}|u|^2dx\right)^{\frac{1}{\tau_3}}\left(m'_3\int_{\{|u|\leq 1\}}[h(|u|^2)]^{2^*}dx\right)^{\frac{1}{\tau_4}}\nonumber\\
&\quad+\left( m_4\int_{\{|u|>1\}}|u|^2dx\right)^{\frac{1}{\tilde{\tau}_3}}\left(m'_4\int_{\{|u|>1\}}[h(|u|^2)]^{2^*}dx\right)^{\frac{1}{\tilde{\tau}_4}}\nonumber\\
&\leq \left( m_3\int_{\mathbb{R}^N}|u|^2dx\right)^{\frac{1}{\tau_3}}\left(m'_3\int_{\mathbb{R}^N}[h(|u|^2)]^{2^*}dx\right)^{\frac{1}{\tau_4}}\nonumber\\
&\quad+\left( m_4\int_{\mathbb{R}^N}|u|^2dx\right)^{\frac{1}{\tilde{\tau}_3}}\left(m'_4\int_{\mathbb{R}^N}[h(|u|^2)]^{2^*}dx\right)^{\frac{1}{\tilde{\tau}_4}}\nonumber\\
&\leq \left( m_3\int_{\mathbb{R}^N}|u_0|^2dx\right)^{\frac{1}{\tau_3}}\left(m'_3 C_s\left(\int_{\mathbb{R}^N}|\nabla h(|u|^2)|^2dx\right)^{\frac{2^*}{2}}\right)^{\frac{1}{\tau_4}}\nonumber\\
&\quad +\left( m_4\int_{\mathbb{R}^N}|u_0|^2dx\right)^{\frac{1}{\tilde{\tau}_3}}\left(m'_4 C_s\left(\int_{\mathbb{R}^N}|\nabla h(|u|^2)|^2dx\right)^{\frac{2^*}{2}}\right)^{\frac{1}{\tilde{\tau}_4}}\nonumber\\
&\leq \left( m_3\|u_0\|^2_{L^2}\right)^{\frac{1}{\tau_3}}\left(m'_3 C_s\right)^{\frac{1}{\tau_4}}\left(\int_{\mathbb{R}^N}|\nabla h(|u|^2)|^2dx\right)^{\frac{2^*}{2\tau_4}}\nonumber\\
&\quad+\left( m_4\|u_0\|^2_{L^2}\right)^{\frac{1}{\tilde{\tau}_3}}\left(m'_4 C_s\right)^{\frac{1}{\tilde{\tau}_4}}\left(\int_{\mathbb{R}^N}|\nabla h(|u|^2)|^2dx\right)^{\frac{2^*}{2\tilde{\tau}_4}}.\label{371}
\end{align}
Here
\begin{align}
\frac{1}{\tau_3}=\frac{\gamma_2-r}{\gamma_2-\gamma_1},\quad \frac{1}{\tau_4}=\frac{r-\gamma_1}{\gamma_2-\gamma_1},\label{36x2}\\
\frac{1}{\tilde{\tau}_3}=\frac{\tilde{\gamma}_2-r}{\tilde{\gamma}_2-\tilde{\gamma}_1},\quad \frac{1}{\tilde{\tau}_4}=\frac{r-\tilde{\gamma}_1}{\tilde{\gamma}_2-\tilde{\gamma}_1}.\label{327x1'}
\end{align}

Noticing that
$$
\int_{\mathbb{R}^N}|\nabla h(|u|^2)|^2dx\leq \int_{\mathbb{R}^N}\Psi(u)dx,
$$
we have
\begin{align}
&\quad\left(\int_0^{+\infty}\left(\int_{\mathbb{R}^N}|G_1(|u|^2)|^rdx\right)^{\frac{q}{r}}dt\right)^{\frac{1}{q}}\nonumber\\
&\leq C(u_0,r,\gamma_1,\gamma_2)\tilde{c}^{\frac{1}{q}}_3\tilde{c}_4\left(\int_0^{+\infty}\left(\int_{\mathbb{R}^N}|\nabla h(|u|^2)|^2dx\right)^{\frac{2^*q}{2r\tau_4}}dt\right)^{\frac{1}{q}}\nonumber\\
&\quad+ \tilde{C}(u_0,r,\tilde{\gamma}_1,\tilde{\gamma}_2)\tilde{c}^{\frac{1}{q}}_3\tilde{c}_4\left(\int_0^{+\infty}\left(\int_{\mathbb{R}^N}|\nabla h(|u|^2)|^2dx\right)^{\frac{2^*q}{2r\tilde{\tau}_4}}dt\right)^{\frac{1}{q}}\nonumber\\
&\leq C_4(u_0,r,\gamma_1,\gamma_2)\left(\int_0^{+\infty}\left(\int_{\mathbb{R}^N}\Psi(u)dx\right)^{\frac{2^*q}{2r\tau_4}}dt\right)^{\frac{1}{q}}\nonumber\\
&\quad+ \tilde{C}_4(u_0,r,\tilde{\gamma}_1,\tilde{\gamma}_2)\left(\int_0^{+\infty}
\left(\int_{\mathbb{R}^N}\Psi(u)dx\right)^{\frac{2^*q}{2r\tilde{\tau}_4}}dt\right)^{\frac{1}{q}}. \label{36x3}
\end{align}
Here  $\tilde{c}_3=1$ if $q\leq r$, $\tilde{c}_3=2^{\frac{q-r}{r}}$ if $q>r$, $\tilde{c}_4=1$ if $q>1$, $\tilde{c}_4=2^{\frac{1-q}{q}}$ if $q\leq 1$,
\begin{align}
C(u_0,r,\gamma_1,\gamma_2)=\left(m_3\|u_0|^2_{L^2}\right)^{\frac{1}{r\tau_3}}\left(m'_3 C_s\right)^{\frac{1}{r\tau_4}},\quad C_4(u_0,r,\gamma_1,\gamma_2)=C(u_0,r,\gamma_1,\gamma_2)\tilde{c}^{\frac{1}{q}}_3\tilde{c}_4\label{327w1}\\
\tilde{C}(u_0,r,\tilde{\gamma}_1,\tilde{\gamma}_2)=\left( m_4\|u_0|^2_{L^2}\right)^{\frac{1}{r\tilde{\tau}_3}}\left(m'_4 C_s\right)^{\frac{1}{r\tilde{\tau}_4}},\quad \tilde{C}_4(u_0,r,\tilde{\gamma}_1,\tilde{\gamma}_2)=\tilde{C}(u_0,r,\tilde{\gamma}_1,\tilde{\gamma}_2)\tilde{c}^{\frac{1}{q}}_3\tilde{c}_4.\label{36x4}
\end{align}

{\bf Case (1).} By (\ref{36x3}), using (\ref{326x3}) and (\ref{326x6}), we get
\begin{align}
&\quad\left(\int_0^{+\infty}\left(\int_{\mathbb{R}^N}|G_1(|u|^2)|^rdx\right)^{\frac{q}{r}}dt\right)^{\frac{1}{q}}\nonumber\\
&\leq C_4(u_0,r,\gamma_1,\gamma_2)\left(\int_0^1[\frac{2E(u_0)[1+M_r(u_0)]}{1-M_r(u_0)}]^{\frac{2^*q}{2r\tau_4}}dt+\int_1^{+\infty} \left(\frac{C(u_0)[1+M_r(u_0)]}{4[1-M_r(u_0)]t^2}\right)^{\frac{2^*q}{2r\tau_4}}dt\right)^{\frac{1}{q}}\nonumber\\
&\quad+\tilde{C}_4(u_0,r,\tilde{\gamma}_1,\tilde{\gamma}_2)\left(\int_0^1[\frac{2E(u_0)[1+M_r(u_0)]}{1-M_r(u_0)}]^{\frac{2^*q}{2r\tilde{\tau}_4}}dt+\int_1^{+\infty} \left(\frac{C(u_0)[1+M_r(u_0)]}{4[1-M_r(u_0)]t^2}\right)^{\frac{2^*q}{2r\tilde{\tau}_4}}dt\right)^{\frac{1}{q}}\nonumber\\
&=C_4(u_0,r,\gamma_1,\gamma_2)\tilde{c}_4\left([\frac{2E(u_0)[1+M_r(u_0)]}{1-M_r(u_0)}]^{\frac{2^*}{2r\tau_4}}
+\left(\frac{C(u_0)[1+M_r(u_0)]}{4[1-M_r(u_0)]}\right)^{\frac{2^*}{2r\tau_4}}\left(\frac{r\tau_4}{2^*q-r\tau_4}\right)^{\frac{1}{q}}\right)\nonumber\\
&\quad+\tilde{C}_4(u_0,r,\gamma_1,\gamma_2)\tilde{c}_4\left([\frac{2E(u_0)[1+M_r(u_0)]}{1-M_r(u_0)}]^{\frac{2^*}{2r\tilde{\tau}_4}}
+\left(\frac{C(u_0)[1+M_r(u_0)]}{4[1-M_r(u_0)]}\right)^{\frac{2^*}{2r\tilde{\tau}_4}}\left(\frac{r\tilde{\tau}_4}{2^*q-r\tilde{\tau}_4}\right)^{\frac{1}{q}}\right).\nonumber\\.\label{37C3}
\end{align}

{\bf Case (2).}  By (\ref{36x3}), using (\ref{326x3}) and (\ref{327x1}), we obtain
\begin{align}
&\quad\left(\int_0^{+\infty}\left(\int_{\mathbb{R}^N}|G_1(|u|^2)|^rdx\right)^{\frac{q}{r}}dt\right)^{\frac{1}{q}}\nonumber\\
&\leq C_5(u_0,r,\gamma_1,\gamma_2)\left\{\left(\int_0^1[2E(u_0)]^{\frac{2^*q}{2r\tau_4}}dt\right)^{\frac{1}{q}}+\right.\nonumber\\
&\qquad\left.\left(\int_1^{+\infty} \left(\frac{C(u_0)}{4t^2}+\frac{4lE(u_0)[1+M_r(u_0)]^2+C(u_0)[1-M_r(u_0)]}{4[1-M_r(u_0)]t^{2-\frac{l[1+M_r(u_0)]}{1-M_r(u_0)}}}
\right)^{\frac{2^*q}{2r\tau_4}}dt \right)^{\frac{1}{q}}\right\}\nonumber\\
&\quad+C_6(u_0,r,\tilde{\gamma}_1,\tilde{\gamma}_2)\left\{\left(\int_0^1[2E(u_0)]^{\frac{2^*q}{2r\tilde{\tau}_4}}dt\right)^{\frac{1}{q}}+\right.\nonumber\\
&\qquad\left.\left(\int_1^{+\infty} \left(\frac{C(u_0)}{4t^2}+\frac{4lE(u_0)[1+M_r(u_0)]^2+C(u_0)[1-M_r(u_0)]}{4[1-M_r(u_0)]
t^{2-\frac{l[1+M_r(u_0)]}{1-M_r(u_0)}}}\right)^{\frac{2^*q}{2r\tilde{\tau}_4}}dt \right)^{\frac{1}{q}}\right\}\nonumber\\
&\leq C_5(u_0,r,\gamma_1,\gamma_2)\left\{[2E(u_0)]^{\frac{2^*}{2r\tau_4}}
+\tilde{C}_1^{\frac{1}{q}}\tilde{c}_4\left(\frac{C(u_0)}{4}\right)^{\frac{2^*}{2r\tau_4}}\left(\frac{r\tau_4}{2^*q-r\tau_4}   \right)^{\frac{1}{q}}\right.\nonumber\\
&\left.+\tilde{C}_1^{\frac{1}{q}}\tilde{c}_4\left(\frac{4lE(u_0)[1+M_r(u_0)]^2+C(u_0)[1-M_r(u_0)]}{4[1-M_r(u_0)]}\right)^{\frac{2^*}{2r\tau_4}}
\left(\int_1^{+\infty} \frac{1}{t^{\frac{2^*q}{2r\tau_4}[2-\frac{l[1+M_r(u_0)]}{1-M_r(u_0)}]}}dt \right)^{\frac{1}{q}}\right\}\nonumber\\
&\quad+C_6(u_0,r,\tilde{\gamma}_1,\tilde{\gamma}_2)\left\{[2E(u_0)]^{\frac{2^*}{2r\tilde{\tau}_4}}
+\tilde{C}_2^{\frac{1}{q}}\tilde{c}_4\left(\frac{C(u_0)}{4}\right)^{\frac{2^*}{2r\tilde{\tau}_4}}\left(\frac{r\tilde{\tau}_4}{2^*q-r\tilde{\tau}_4}   \right)^{\frac{1}{q}}\right.\nonumber\\
&\left.+\tilde{C}_2^{\frac{1}{q}}\tilde{c}_4\left(\frac{4lE(u_0)[1+M_r(u_0)]^2+C(u_0)[1-M_r(u_0)]}{4[1-M_r(u_0)]}\right)^{\frac{2^*}{2r\tilde{\tau}_4}}
\left(\int_1^{+\infty} \frac{1}{t^{\frac{2^*q}{2r\tilde{\tau}_4}[2-\frac{l[1+M_r(u_0)]}{1-M_r(u_0)}]}}dt \right)^{\frac{1}{q}}\right\}\nonumber\\
&=C_5(u_0,r,\gamma_1,\gamma_2)\left([2E(u_0)]^{\frac{2^*}{2r\tau_4}}+\tilde{C}_1^{\frac{1}{q}}\tilde{c}_4\left(\frac{C(u_0)}{4}\right)^{\frac{2^*}{2r\tau_4}}
\left(\frac{r\tau_4}{2^*q-r\tau_4}   \right)^{\frac{1}{q}}\right)\nonumber\\
&\quad+C_5(u_0,r,\gamma_1,\gamma_2)\tilde{C}_1^{\frac{1}{q}}\tilde{c}_4
\left(\frac{4lE(u_0)[1+M_r(u_0)]^2+C(u_0)[1-M_r(u_0)]}{4[1-M_r(u_0)]}\right)^{\frac{2^*}{2r\tau_4}}\nonumber\\
&\qquad \times \left(\frac{2r\tau_4[1-M_r(u_0)]}{(22^*q-2r\tau_4)[1-M_r(u_0)]-2^*ql[1+M_r(u_0)]} \right)^{\frac{1}{q}}\nonumber\\
&+\quad C_6(u_0,r,\tilde{\gamma}_1,\tilde{\gamma}_2)\left([2E(u_0)]^{\frac{2^*}{2r\tilde{\tau}_4}}
+\tilde{C}_2^{\frac{1}{q}}\tilde{c}_4\left(\frac{C(u_0)}{4}\right)^{\frac{2^*}{2r\tilde{\tau}_4}}\left(\frac{r\tilde{\tau}_4}{2^*q-r\tilde{\tau}_4}   \right)^{\frac{1}{q}}\right)\nonumber\\
&\quad+C_6(u_0,r,\tilde{\gamma}_1,\tilde{\gamma}_2)\tilde{C}_2^{\frac{1}{q}}\tilde{c}_4
\left(\frac{4lE(u_0)[1+M_r(u_0)]^2+C(u_0)[1-M_r(u_0)]}{4[1-M_r(u_0)]}\right)^{\frac{2^*}{2r\tilde{\tau}_4}}\nonumber\\
&\qquad \times \left(\frac{2r\tilde{\tau}_4[1-M_r(u_0)]}{(22^*q-2r\tilde{\tau}_4)[1-M_r(u_0)]-2^*ql[1+M_r(u_0)]}\right)^{\frac{1}{q}}.\label{37C3}
\end{align}
Here $\tilde{C}_1=1$ if $2^*q\leq 2r\tau_4$, $\tilde{C}_1=2^{\frac{2^*q-2r\tau_4}{2r\tau_4}}$ if $2^*q>2r\tau_4$; $\tilde{C}_2=1$ if $2^*q\leq 2r\tilde{\tau}_4$, $\tilde{C}_2=2^{\frac{2^*q-2r\tilde{\tau}_4}{2r\tilde{\tau}_4}}$ if $2^*q>2r\tilde{\tau}_4$. And
\begin{align}
C_5(u_0,r,\gamma_1,\gamma_2)=C_4(u_0,r,\gamma_1,\gamma_2)\tilde{c}_4\left(\frac{1+M_r(u_0)}{1-M_r(u_0)}\right)^{\frac{2^*}{2r\tau_4}},\\
C_6(u_0,r,\tilde{\gamma}_1,\tilde{\gamma}_2)=\tilde{C}_4(u_0,r,\tilde{\gamma}_1,\tilde{\gamma}_2)\tilde{c}_4
\left(\frac{1+M_r(u_0)}{1-M_r(u_0)}\right)^{\frac{2^*}{2r\tilde{\tau}_4}}.
\end{align}

To illustrate our results, we give some examples of $h(s)$ and $F(s)$ below.

{\bf Example 5.1.} If $F(s)\equiv 0$ or $F(s)=-s^q$, $q>0$, then the solution of (\ref{1}A) is global existence. Morawetz estimates and space bounds for the solution can be established, for example,
$$
\int_0^{+\infty}\int_{\mathbb{R}^N}[|\nabla h(|u|^2)|^2+A[h(|u|^2)]^{2^*}]dxdt\leq C,
$$
or
$$
\int_0^{+\infty}\int_{\mathbb{R}^N}[|\nabla h(|u|^2)|^2+|u|^{2q+2}+A[h(|u|^2)]^{2^*}]dxdt\leq C.
$$

{\bf Example 5.2.} If $h(s)=s^{\alpha}$, $\alpha\geq \frac{1}{2}$, $F(s)=s^{\tilde{p}}$, $0<\tilde{p}<2\alpha\cdot 2^*-1$, then the solution of (\ref{1}A) is global existence.
Especially, if $\tilde{p}=2\alpha-1+\frac{2}{N}$, then (\ref{6262}) and (\ref{6263}) are satisfied. We can get Morawetz estimates and space bounds for the solution
if initial data $u_0$ satisfies (\ref{6261}), for example,
$$
\||u|^{4\alpha+\frac{4}{N}}\|_{L^q(\mathbb{R}^+)L^r(\mathbb{R}^N)}=\left(\int_0^{\infty}\left(\int_{\mathbb{R}^N}[|u|^{4\alpha+\frac{4}{N}}]^rdx\right)^{\frac{q}{r}}dt\right)^{\frac{1}{q}}
\leq C
$$
for suitable $q$ and $r$.

{\bf Example 5.3.} If $h(s)=a_1s^{\alpha_1}+...+a_ms^{\alpha_m}$, $$F(s)=b_1s^{p_1}+...+b_ns^{p_n}-c_1s^{q_1}-...-c_rs^{q_r},$$ the coefficients $a_1$, ..., $a_m$ $b_1$, ..., $b_n$, $c_1$, ...., $c_r$ are positive, $0<\alpha_1<..<\alpha_m$, $\alpha_m\geq \frac{1}{2}$,
$0<p_1<...<p_n$, $0<q_1<...<q_r$, $p_1=2\alpha_1-1+\frac{2}{N}$, $p_n=2\alpha_m-1+\frac{2}{N}$, then (\ref{6262}) and (\ref{6263}) are satisfied. We can get Morawetz estimates and space bounds for the solution
if initial data $u_0$ satisfies (\ref{6261}), for example,
\begin{align*}
&\quad\int_0^{\infty}\left(\int_{\mathbb{R}^N}|\nabla [a_1|u|^{2\alpha_1}+...+a_m|u|^{2\alpha_m}]|^2+A[a_1|u|^{2\alpha_1}+...+a_m|u|^{2\alpha_m}]^{2^*}dx\right)^pdt\nonumber\\
&+\int_0^{\infty}\left(\int_{\mathbb{R}^N} [b_1|u|^{2p_1+2}+...+b_n|u|^{2p_n+2}]+[c_1|u|^{2q_1+2}+...+c_r|u|^{2q_r+2}]dx\right)^pdt\nonumber\\
&\leq C
\end{align*}
for suitable $p$.

{\bf Remark 5.1.} Under the assumption on $h(s)$ in (\ref{3251}), the following model is the special case of (\ref{1}) with $F(|u|^2)u=a|u|^{2^*-2}u$
\begin{equation}
\label{420w1} \left\{
\begin{array}{lll}
iu_t=\Delta u+2uh'(|u|^2)\Delta h(|u|^2)+a|u|^{2^*-2}u\mp A[h(|u|^2]^{2^*-1}h'(|u|^2)u,\ x\in \mathbb{R}^N, \ t>0\\
u(x,0)=u_0(x),\quad x\in \mathbb{R}^N.
\end{array}\right.
\end{equation}

If $h(s)=as^{\frac{1}{2}}$, then (\ref{1}) becomes
\begin{equation}
\label{421w1} \left\{
\begin{array}{lll}
iu_t=\Delta u+\frac{u}{|u|}\Delta |u|^2+b|u|^{2^*-2}u,\ x\in \mathbb{R}^N, \ t>0\\
u(x,0)=u_0(x),\quad x\in \mathbb{R}^N.
\end{array}\right.
\end{equation}

Naturally, the corresponding results on (\ref{1}) hold in the two special cases.

In the last part of this section, we would like to compare the results on (\ref{1}) to those on (\ref{1'}).

{\bf Remark 5.2.} (1). Since $A>0$, the results about the conditions on the global existence of the solution to (\ref{1}A) and blowup of (\ref{1}B) in this paper are differ from those on (\ref{1'}) in \cite{SongWang1}, they cannot be covered each other.

(2). However, mass, energy and the pseudoconformal conservation laws for the global solution of (\ref{1'}) are similar to these for the global solution of (\ref{1}A). If we look (\ref{1'}) as the special case of (\ref{1}) with $A=0$, these conservation laws for (\ref{1}A) can cover those for (\ref{1'}). Although we didn't establish Morawetz estimates and spacetime bounds for the global solution of (\ref{1'}) in \cite{SongWang1}), we can prove the corresponding Morawetz estimates and spacetime bounds for the global solution of (\ref{1'}) by letting $A=0$ in these for (\ref{1}) under the same assumptions on $h(s)$, $F_1(s)$ and $F_2(s)$. For example, the corresponding result on problem (\ref{1'}) to (\ref{2211'}) is
\begin{align}
\int_0^{\infty}\int_{\mathbb{R}^N}\frac{\left[|\nabla h(|u|^2)|^2+|G_1(|u|^2)|+|G_2(|u|^2)|\right]^{\theta}}{n_1(x,t)}dxdt\leq M'_1(u_0,\theta),
\end{align}
the corresponding result on problem (\ref{1'}) to (\ref{37w1}) is
\begin{align}
\left(\int_0^{\infty}\left(\int_{\mathbb{R}^N}\left[|\nabla h(|u|^2)|^2+|G_1(|u|^2)|+|G_2(|u|^2)|\right]dx\right)^pdt\right)^{\frac{1}{p}}\leq C'(u_0,p).
\end{align}

\end{document}